\documentclass[a4paper, 10pt]{article}
\usepackage[utf8]{inputenc}
\usepackage{jheparxiv}
\usepackage{amsmath,mathrsfs,bm}
\usepackage[final]{showkeys}
\usepackage{draft}
\usepackage[normalem]{ulem}

\newcommand{\IC}[0]{\mathbb{C}}
\newcommand{\IR}[0]{\mathbb{R}}
\newcommand*{\temp}{\multicolumn{1}{r|}{}}
\renewcommand{\L}{\ell}
\renewcommand{\R}{r}
\newcommand{\CM}{\overline{M}}
\newcommand{\CML}{\overline{M}_\L}
\newcommand{\CMR}{\overline{M}_\R}
\renewcommand{\I}{\mathfrak{I}}

\title{Conformal boundaries of Minkowski superspace \\and their super cuts}

\author{Nicolas Boulanger$^{\dagger}$, Yannick Herfray$^{\ddagger}$,
Noémie Parrini$^{\dagger}$\footnote{Research Fellow of the F.R.S.-FNRS (Belgium).}}

\affiliation{$^{\dagger}$Physique de l'Univers, Champs et Gravitation, 
Universit\'e de Mons -- UMONS,\\Place du Parc 20, 7000 Mons, Belgium}

\affiliation{$^{\ddagger}$Institut Denis Poisson UMR 7013, Université de Tours, 
Parc de Grandmont, 37200 Tours, France}

\emailAdd{nicolas.boulanger@umons.ac.be, yannick.herfray@univ-tours.fr, noemie.parrini@umons.ac.be}

 \abstract{In this article we carry out a detailed investigation of the geometric nature 
 of the points at infinity of Minkowski superspace. 
 It turns out that there are several sets of points forming the superconformal boundary of 
 Minkowski superspace: on top of a well-behaved super $\mathscr{I}$, 
 we find other sets that we exhibit and study.
 We also study the intersection of these boundaries with super null cones 
 and explicitly construct the corresponding space of super cuts.}

\begin{document}

\maketitle

\section{Introduction}

Penrose's conformal compactification of spacetimes \cite{penrose_asymptotic_1963} is an essential tool in general relativity \cite{geroch_asymptotic_1977,wald_general_1984,frauendiener_conformal_2004,ashtekar_geometry_2015}. It was introduced as mean to both geometrize and give a global meaning to Bondi--Van Der Burg--Metzner--Sachs notion of asymptotically flat spacetimes \cite{bondi_gravitational_1962,sachs_gravitational_1962}. 
The very notion of conformal boundary is also at the basis 
of the AdS/CFT correspondence \cite{witten_anti_1998,maldacena_large_1998}.

In the context of supergeometry it is well known that there exists a notion of superconformal compactification of Minkowski superspace. The construction is detailed in Manin's book \cite{manin_gauge_1988} and, as we will review, is most naturally realised in terms of supertwistors geometry \cite{ferber_supertwistors_1978}. 
The resulting conformally compactified Minkowski superspace 
$\overline{M}^{4|4 \mathcal{N}}$ naturally is a homogeneous space\footnote{Realisations of Minkowski and AdS superspaces as coset spaces is a classical topics in supersymmetry, we refer to the monographs \cite{manin_gauge_1988,castellani_supergravity_1991,buchbinder_ideas_1998,galperin_harmonic_2001} for further details and references and to \cite{figueroa-ofarrill_kinematical_2019} for the classification of all $\mathcal{N}=1$ kinematical superspaces. In this article, we will focus on the asymptotic boundaries of these models.} 
for the super conformal group $\text{SU}(2,2|\mathcal{N})$:
\begin{equation}\label{Homogeneous space: intro,  compactified  real Minkowski}
    \overline{M}^{4|4 \mathcal{N}} = \frac{\text{SU}(2,2|\mathcal{N})}{(\mathbb{R^*}\times \text{SL}(2,\mathbb{\IC}) \times \text{SU}(\mathcal{N})) \ltimes \mathcal{T}^{4|4 \mathcal{N}}}\;.
\end{equation}
AdS superspace $AdS^{4|4\mathcal{N}} \hookrightarrow \overline{M}^{4|4 \mathcal{N}}$ 
is recovered by breaking superconformal invariance down to 
$\text{OSp}(\mathcal{N}|4) \subset \text{SU}(2,2|\mathcal{N})$, 
the supergroup of isometries of AdS superspace. 
The action of $\text{OSp}(\mathcal{N}|4)$ on conformally compactified 
Minowski superspace $\overline{M}^{4|4\mathcal{N}}$ is not transitive and resulting orbits
include AdS superspace $AdS^{4|4\mathcal{N}} \hookrightarrow\overline{M}^{4|4\mathcal{N}}$ and its conformal boundary $\overline{M}^{3|2\mathcal{N}} \hookrightarrow\overline{M}^{4|4\mathcal{N}}$:
\begin{align}\label{Homogeneous space: intro, super AdS}
    AdS^{4|4\mathcal{N}} & \simeq  \frac{\text{OSp}(\mathcal{N}|4)}{\text{SL}(2,\IC) \times \text{SO}(\mathcal{N})}\;,&
    \overline{M}^{3|2\mathcal{N}} & \simeq \frac{\text{OSp}(\mathcal{N}|4)}{(\mathbb{R}^* \times \text{SL}(2,\mathbb{R}) \times \text{SO}(\mathcal{N}))\ltimes \mathcal{T}^{3\vert 2\mathcal{N}}}\;.
\end{align}
Both supermanifolds naturally fit inside the bigger supermanifold $\overline{M}^{4|4\mathcal{N}}$. 
This superembeding construction, as well as its extended versions, as been studied with a lot of care in \cite{kuzenko_supertwistor_2021,koning_embedding_2023,kuzenko_embedding_2023}. 

Similarly, as has been used in numerous context \cite{kotrla_supertwistors_1985,lukierski_general_1988,howe_harmonic_1994,hartwell_n_1995,howe_superspace_1995,kuzenko_compactified_2006,wolf_supertwistor_2006,boels_supersymmetric_2007,wolf_first_2010,kuzenko_off-shell_2011,kuzenko_conformally_2012,adamo_twistor_2013,buchbinder_superconformal_2015}, 
by breaking superconformal invariance down to the super Poincaré group 
$\text{ISO}(3,1|\mathcal{N}) \subset \text{SU}(2,2|\mathcal{N})$ one obtains 
an embedding 
$M^{4|4\mathcal{N}} \hookrightarrow \overline{M}^{4|4\mathcal{N}}$ 
of Minkowski superspace. 
However, it seems that not so much interest has been given in the literature to the 
investigation of the nature of the boundary ``at infinity'' of Minkowski superspace. 
In the first two sections of this work we investigate this aspect in details. 
As a result, we find the following invariant decomposition 
of conformally compactified Minowski superspace
under the action of the 
super Poincaré group\footnote{In this introduction we focus on real Minkowski  
superspace, however the complexifed superconformal compactifiation appears most naturally 
in this context. This is of interest because it makes the underlying chirality 
properties of Minkowski superspace manifest. 
All these aspects will be discussed in details in the body of the paper.}:

\begin{equation}
    \begin{array}{ccccccccccc}
       \overline{M}^{4|4\mathcal{N}}  &=&  M^{4|4\mathcal{N}} & \bigsqcup & \mathscr{I}^{3|2\mathcal{N}} & \bigsqcup & \mathcal{H}^{} & \bigsqcup &  \I  & \bigsqcup &\iota 
    \end{array}\;.
\end{equation}
The first supermanifold appearing above obviously is Minkowski superspace,
\begin{align}\label{Homogeneous space: intro, real Minkowski}
     M^{4|4\mathcal{N}}& \simeq  \frac{\text{ISO}(1,3|\mathcal{N})}{\text{SL}(2,\IC) \times \text{SU}(\mathcal{N})}\;,
\end{align}
while the four remaining subspaces
are four other sets of points left invariant by the action of the super Poincaré group. 
In the core of this paper we will give explicit parametrisations 
of each of these spaces ``at infinity'' of Minkowski superspace. 
We here briefly sum up the resulting picture: the most well-behaved is super null infinity. 
It is an homogeneous superspace\footnote{Even though this realisation of super 
null infinity is certainly known to experts, we could not find any reference 
on the subject. See however {\cite{adamo_perturbative_2014}} for the use of null 
infinity in a superspace context.},
\begin{align}\label{Homogeneous space: intro, real scri}
    \mathscr{I}^{3|2\mathcal{N}} & \simeq \frac{\text{ISO}(1,3|\mathcal{N})}{(\mathbb{R}^* \times \text{ISO}(2) \times \text{SU}(\mathcal{N}) )\ltimes \mathcal{T}^{3|2\mathcal{N}}}\;,
\end{align}
and as such always is a supermanifold. On the other other hand, the set of points 
that we note $\mathcal{H}$ and $\I$ only are supermanifolds in the chiral 
(left or right) complexified setting, 
in which case they respectively have complex dimensions $(3|2\cN)$ 
and $(0|2 \cN)$. For comparison, complexified chiral (left or right) Minkowski 
and null infinity respectively have complex dimensions $(4|2\cN)$ and $(3|\cN)$, 
thus $\mathcal{H}$ has co-dimension $(1|0)$.
However they are never homogeneous spaces: the action of the 
super Poincaré group happens not to be transitive. 
If one tries, as we shall explain in the 
bulk of the paper, to restrict to orbits of the super Poincaré group, 
the resulting sets turn out not to  be 
supermanifolds. This is indeed a peculiar  
phenomenon of supergeometries 
that not all orbits of a supergroup are supermanifolds. 
The reason is that the stabilizer of a point in the orbit may not be a supergroup. 
The final invariant set $\iota$ is just a point: it stands for both time-like and 
space-like infinity in the conformal compactification. 

To further elucidate the nature of these boundaries, we turn to different 
but closely related geometrical objects of Minkowski superspace: super null rays. 
These are natural supermanifolds of dimension {$(1|2{\cal N})$} 
which generalise null geodesics \cite{manin_gauge_1988}: if 
$(x_0^{AA'}, \theta_0^{IB'})$ 
is a point in Minkowski superspace then the super-null ray in the direction 
$[\pi^{A'}] \in \IC P^1$ is given by the parametric equations
\begin{align}
x^{AA'} &= x_0^{AA'} + \frac{1}{2}\left( \pi^{A'} \bar{\theta}_0{}_I{}^A \epsilon^I 
+ \theta^{IA'}_0 \bar{\pi}^A \bar{\epsilon}_I \right)  +\epsilon \,\pi^{A'}\bar{\pi}^{A}\;,
\\
\theta^{IB'} &= \theta_0^{IB'} + \epsilon^I\pi^{B'}\;,
\end{align}
where $(\epsilon,\epsilon^I) \in \mathbb{R}^{(1|0)}\times \IC^{0|\cN}$ 
are coordinates 
along the super null ray. In this way, each point of Minkowski superspace is 
associated 
with a Riemann sphere $\IC P^1$ of super null rays passing through it. 
By making use of the fact that super null rays are points in the 
super ambitwistor space
\cite{Witten:1985nt,manin_gauge_1988,Mason:2013sva}, 
we shall see that super null cones intersect the different boundaries at infinity of 
Minkowski superspace along preferred super cuts. 
We will obtain explicit expressions for the resulting space of cuts.

Finally, one can reverse the logic to gain some insight about the boundaries at infinity: 
in the bosonic geometry, null infinity $\mathscr{I}$ can be understood as a null cone 
emanating from time-like infinity $\iota\,$. This interpretation is possible because null 
cones are conformal invariants. Similarly, in the supergeometric context we can use the fact 
that super null rays are superconformal invariant to interpret points at infinity: it turns 
out that while super $\mathscr{I}$ genuinely is the super null cone emanating from the point 
$\iota$, the boundary $\mathcal{H}$ is the union of all super null rays emanating from the 
purely fermionic points $\I\,$. Here $\I$ should be thought of 
as the results of translations of $\iota$ in fermionic directions.

The paper is organised as follows: to be self-contained and 
present our notations, we first review how to explicitly 
realise the superconformal compactification of Minkowski space in terms of supertwistors. 
We then discuss the different sets of points invariant under the action of the 
super Poincaré group. Finally, we recall the equivalence between ambitwistors and 
super null rays and make use of this identification to construct cuts along super null 
infinity and the other boundaries.

\section{Superconformal compactification of Minkowski superspace}
\label{sec:Minkowski superspace}

In this section we review the well-known superconformal compactification 
of Minkoswki superspace in terms of supertwistors \cite{ferber_supertwistors_1978,manin_gauge_1988}. 
We also review the real forms of the complex superconformal group and algebra.

\subsection{Supertwistors}

\subsubsection{Supertwistor space}\label{subsubsec: super twistor space}
 
We take the supertwistor space to be the fundamental representation of the complexified 
superconformal group $\text{SL}(4 \vert \cN  )$.
In coordinates, a supertwistor $Z^{\hat\alpha}$ will be represented by an element of $\IC^{4\vert \cN}$:
\begin{equation} Z^{\hat\alpha} = 
    \begin{pmatrix}
        \omega^A \\
        \pi_{A'} \\
        \theta^I
    \end{pmatrix},
\end{equation}
where
\begin{align}
\omega^A &= \begin{pmatrix}
    \omega^0 \\
    \omega^1
\end{pmatrix} \in \IC_c^2\;,
&
\pi_{A'} &= \begin{pmatrix}
    \pi_{0'} \\
    \pi_{1'} 
\end{pmatrix} \in \IC^2_c\;, 
& 
 \theta^I = \begin{pmatrix}
     \theta^1 \\
     \vdots \\
     \theta^{\cN}
 \end{pmatrix}  \in \IC ^{\cN}_a\;.
\end{align}

The supergroup $\text{SL}(4 \vert \cN)$ is parametrised by  
\begin{equation}\label{eqn: SL4} X^{\hat{\alpha}}{}_{\hat{\beta}}=
    \begin{pmatrix}
		{M^A}_B & i\,T^{AB'} & {Q^A}_J \\
        -i\,K_{A'B} & -{\widetilde{M}}_{A'}{}^{B'} & S_{A'J} \\
        {\widetilde{S}}^I{}_B & \widetilde{Q}^{IB'} & R^I{}_J
	\end{pmatrix}\;,
\end{equation}
where 
\begin{align}
     \begin{pmatrix}
         {M^A}_B & iT^{AB'} \\
         -iK_{A'B} & -{\widetilde{M}}_{A'}{}^{B'}
     \end{pmatrix} &\in \text{GL}(4, \IC_c)\;,
&
R^I{}_J &\in \text{GL}(\cN, \IC_c)\;,
& \text{Ber}(X) = 1\;,
\end{align}
and ${Q^A}_J\,$, $\widetilde{Q}^{IB'}\,$, $S_{A'J}\,$, 
and ${\widetilde{S}}^I{}_B$ are elements of $\IC_a^{2\cN}\,$.
The supergroup $\text{SL}(4 \vert \cN)$ acts linearly on the supertwistors:
\begin{align}
    Z^{\hat\alpha} \mapsto X^{\hat\alpha}{}_{\hat\beta}\,Z^{\hat\beta}\;.
\end{align}

\subsubsection{Real form of the superconformal group}

If we restrict to the real case, 
i.e., if we ask to preserve the hermitian form
\begin{align}\label{Hermitian form contraction}
	\overline{Z}^{\overline{\hat{\alpha}}}\, 
    h_{\overline{\hat{\alpha}}\hat{\beta}}  Z^{\hat{\beta}}   
    &=  \overline{\pi}_A\, \omega^A  
    + \overline{\omega}^{A'}\,\pi_{A'} 
    - {\overline{\theta}}^J\, \theta^I \delta{}_{IJ}\;, 
 \\ \mbox{where}\quad h_{\overline{\hat{\alpha}}\hat{\beta}} &= 
	\begin{pmatrix}
		0 & \mathbb{I}_2 & 0 \\
		\mathbb{I}_2 & 0 & 0 \\
		0 & 0 & -\delta{}_{IJ}
	\end{pmatrix}\;,
 \label{hermitianForm}
\end{align}
then the action of $\text{SL}(4\vert \mathcal{N})$ reduces 
to the action of the \emph{real} superconformal group $\text{SU}(2,2|\cal{N})\,$.
A generic element $x$ of $\mathfrak{su}(2,2|\cal{N})$, the subalgebra of matrices $x$ in 
$\mathfrak{sl}(4\vert \mathcal{N})$ such that $h x + x^\dag h = 0$, can be written as 
\begin{equation}
x=    \begin{pmatrix}
        m^A{}_B & i t^{AB'} & q^A{}_J \\
        -i k_{A'B}  &-{\overline{m}}^{B'}{}_{A'}& \overline{s}_{A'J} \\ 
        s^I{}_{B} & \overline{q}^{IB'}  & r^I{}_J
    \end{pmatrix} + c\,A\;,\quad c\in \mathbb{R}_c\;,
\end{equation}
where Tr$\,m =$ Tr$\,m^\dagger\,$,
$t^\dagger = t\,$, $k^\dagger = k$ and $r^\dagger = r\,$, Tr$\,r=0\,$.
The ${\cal N}\times {\cal N}$ matrix $(r^I{}_J)$ belongs to the 
$\mathfrak{su}({\cal N})$ subalgebra while the matrix $A$ is generator of 
$U(1)\,$.
In the case ${\cal N} = 1$, 
the matrix $A$ is given in appendix \ref{appendixSuperConf} to which 
we refer for more details on the superconformal algebra and its presentation.
If one denotes by $\alpha$ and $\beta$ the diagonal elements of the 
matrix $m\,$, the trace constraint on $m$ and $m^\dagger$ implies that 
$\Im(\alpha+\beta)=0\,$, i.e., Tr$\,m\in\mathbb{R}_c$.
Note that 
${m^\dagger}{}_{A'}{}^{B'}=(m^B{}_A)^* =: \overline{m}^{B'}{}_{A'}\,$.

\subsection{Superconformal compactification}

\subsubsection{Complex case}\label{subsec: complex case}

Following \cite{manin_gauge_1988}, the model for full (i.e., non-chiral) 
$\mathcal{N}$ extended compactified \emph{complexified} Minkowski superspace $\overline{M} $ is taken to be the flag manifold $F(2\vert 0, 2\vert \mathcal{N}, \IC^{4\vert \mathcal{N}})$ in the supertwistor space.

There are then two canonical projections from compactified Minkowski superspace to the left/right chiral superspaces: 
\begin{center}\label{left/right chiral superspaces projection}
\begin{tikzpicture}
\node {$\CM = F(2\vert 0, 2\vert \mathcal{N}, \IC^{4\vert \mathcal{N}})$}[sibling distance = 4.5cm]
    child {node {$\CML = Gr(2\vert 0, \IC^{4\vert \mathcal{N}})$ } edge from parent node [left] {$\pi_{\L}$}}
    child {node {$\CMR = Gr(2\vert \mathcal{N}, \IC^{4\vert \mathcal{N}})$} edge from parent node [right] {$\pi_{\R}$}};
\end{tikzpicture}	
\end{center}
explicitly given by
\begin{align*}
    \pi_{\L} : F(2\vert 0, 2\vert \mathcal{N}, \IC^{4\vert \mathcal{N}}) & \longrightarrow Gr(2\vert 0, \IC^{4\vert \mathcal{N}})
    & \pi_{\R} : F(2\vert 0, 2\vert \mathcal{N}, \IC^{4\vert \mathcal{N}}) & \longrightarrow Gr(2\vert \cN, \IC^{4\vert \mathcal{N}})
    \\
    (P_1, P_2) &\longmapsto P_1 
    &
    (P_1, P_2) &\longmapsto P_2\;.
\end{align*}
As we will recall, $\CML$, $\CMR$ and $\CM$ are homogeneous (super)spaces for the complex superconformal group and are respectively of dimensions
\begin{align}
	\dim _{\IC} \CML = \dim _{\IC} \CMR &= 4 \vert 2\mathcal{N},\ & \dim _{\IC} \CM &= 4\vert 4 \mathcal{N}.
\end{align}

We will use grassmanian coordinates $Z^{\hat\alpha b}$ with, 
$\hat\alpha\in  \{{}^A, {}_{A'}, {}^I\}$, $b\in \{1,2\}$ 
to parametrise $Gr(2\vert 0, \IC^{4\vert \mathcal{N}})$ 
(with similar notations for the other grassmanians). 
We call elements $Z^{\hat\alpha b} \in \CML$ bi-supertwistors. 
The elements of $\CMR$ will be denoted by $Z^{\hat\alpha \textbf{c}}$, with $\textbf{c} \in \{1, \cdots , \mathcal{N}+2\}$.
Making use of the isomorphism \begin{align}\label{eqn: grassm duality}
	Gr(2\vert \mathcal{N}, \IC^{4\vert \mathcal{N}}) 
 &\simeq  Gr\big(2\vert 0, {( \IC^{4\vert \mathcal{N} } )}^*\big)\;, 
\end{align}
we equivalently represent elements of $\CMR$ by dual bi-supertwistors 
$\widetilde{Z}_{\hat\alpha}{}^b\,$. 
Elements of $\CM= F(2\vert 0, 2\vert \mathcal{N}, \IC^{4\vert \mathcal{N}})$ 
are then obtained as pairs $(Z^{\hat\alpha b}, {\widetilde{Z}}_{\hat \alpha}{}^c)$ 
of bi-supertwistors and dual bi-supertwistors satisfying 
${\widetilde{Z}}_{\hat\alpha}{}^b Z^{\hat\alpha c} = 0\,$. 
With these notations the projections \eqref{left/right chiral superspaces projection} 
are simply realised as
\begin{center}
\begin{tikzpicture}
\node {$(Z^{\hat\alpha b}, {\widetilde{Z}}_{\hat \alpha}{}^c) \in \CM$}[sibling distance = 4.5cm]
    child {node { $Z^{\hat\alpha b} \in \CML$ } edge from parent node [left] {$\pi_{\L}$}}
    child {node {${\widetilde{Z}}_{\hat \alpha}{}^c \in \CMR$\ \ .} edge from parent node [right] {$\pi_{\R}$}};
\end{tikzpicture}	
\end{center}

The supergroup $\text{SL}(4\vert \mathcal{N})$ acts on $\CML$, $\CMR$ and $\CM$ via the induced action from $\IC^{4\vert \mathcal{N}}$:
\begin{align}\label{eqn: action}
\begin{array}{ccc}
    Z^{\hat{\alpha} b} \in \CML   & \mapsto & 
    X^{\hat\alpha}{}_{\hat\beta} Z^{\hat{\beta} b} \;,\\
    {\widetilde{Z}}_{\hat{\alpha}}{}^b \in \CMR   & \mapsto & 
    {\widetilde{Z}}_{\hat{\beta}}{}^b\,{(X^{-1})}^{\hat\beta}{}_{\hat\alpha}\;, \\    
    ( Z^{\hat{\alpha} b} , {\widetilde{Z}}_{\hat{\alpha}} {}^c )    \in \CM   & \mapsto &  ( X^{\hat\alpha}{}_{\hat\beta} Z^{\hat{\beta} b} ,   
     {\widetilde{Z}}_{\hat{\beta}}{}^c\, (X^{-1})^{\hat\beta}{}_{\hat\alpha})\;.   
\end{array}
\end{align}
This action is transitive in all three cases, which makes $\CM$, $\CML$ and $\CMR$ all 
homogeneous spaces for the complex superconformal group. 
The respective stabilisers are obtained by choosing specific points.

\begin{itemize}
	\item For $\CML\,$, we require \eqref{eqn: SL4} to stabilise the 
    origin bi-supertwistor
	\begin{equation}\label{eqn: origin twistor}
		Y^{\hat\alpha b} := 
		\begin{bmatrix}
			0^{Ab} \\
			\delta_{A'}{}^b \\
			0^{Ib}
		\end{bmatrix}\;,
	\end{equation} 
 where the notation $[\cdot]$ in the above definition is used to denote 
 the equivalence class of elements in $Gr\big(2\vert 0, { \IC^{4\vert \mathcal{N} }}\big)$, for which a representative is denoted between the square brackets. 
	The stabiliser is the set of supermatrices of $\text{SL}(4,\cN)$ of the form
	\begin{equation}\label{eqn: chiral poincare group 1}
		\text{P}_0^\L = \begin{pmatrix}
			{M^A}_B & 0^{AB'} & {Q^A}_J \\
        -i K_{A'B} & -{\widetilde{M}}_{A'}{}^{B'} & S_{A'J} \\
        {\tilde{S}}^I{}_B & 0^{IB'} & R^I{}_J
		\end{pmatrix}\;,
	\end{equation} 
 and forms a parabolic subgroup. 
Therefore, we have the homogeneous space
	\begin{equation}
		\CML \simeq \frac{\text{SL}(4\vert \mathcal{N})}{\text{P}_0^\L}\;.
	\end{equation}
	\item For $\CMR$, we require \eqref{eqn: SL4} 
 to stabilise the dual origin bi-supertwistor
 \begin{equation}\label{eqn: dual origin twistor}
		\widetilde{Y}_{\hat\alpha}{}^b := 
		\begin{bmatrix}
			\delta_{A}{}^b & 0^{A'}{}^b & 0_{I}{}^b
		\end{bmatrix} \;.
\end{equation}
An equivalent way to find the stabiliser of the dual origin bi-supertwistor 
is to use the isomorphism \eqref{eqn: grassm duality} on the following 
$({\mathcal N}+2)$-supertwistor :
Decomposing the index $\textbf{c}$ into $(c,\tt{C})$ with $c \in \{1,2\}$ and 
$\tt{C}$ $\in \{ 1, \cdots , \mathcal{N}\}$, we obtain
\begin{align}
    Y^{\hat\alpha \textbf{c}} 
    = \begin{bmatrix}
    	0^{Ac} & 0^{A\tt{C}} \\
    	\delta_{A'}{}^c & 0_{A'}{}^{\tt{C}} \\
    	0^{Ic} & \delta^{I\tt{C}}
    \end{bmatrix}
\end{align}
as the dual origin bi-supertwistor. The stabiliser is then the set of supermatrices
	\begin{equation}\label{eqn: chiral poincare group 2}
		\text{P}_0^\R = \begin{pmatrix}
			{M^A}_B & 0^{AB'} & {0^A}_J \\
        -i K_{A'B} & -{{\widetilde{M}}_{A'}}{}^{B'} & S_{A'J} \\
        {\tilde{S}}^I{}_B & {\tilde{Q}}^{IB'} & R^I{}_J
		\end{pmatrix}\;,
	\end{equation}	
forming again a parabolic subgroup of $\text{SL}(4\vert \mathcal{N})\,$ and yielding the homogeneous space
	\begin{equation}
		\CMR \simeq \frac{\text{SL}(4\vert \mathcal{N})}{\text{P}_0^\R}\;.
	\end{equation}
 \end{itemize}

 \begin{itemize}
\item For $\CM$, the subgroup preserving the origin 
$(Y^{\hat\alpha b},\widetilde{Y}_{\hat\alpha}{}^b) \in \CM$ is
parametrised by 
	\begin{equation}\label{eqn: poincare group}
		\text{P}_0 = \begin{pmatrix}
			{M^A}_B & 0^{AB'} & {0^A}_J \\
        -i K_{A'B} & -{{\widetilde{M}}_{A'}}{}^{B'} & S_{A'J} \\
        {\tilde{S}}^I{}_B & {0}^{IB'} & R^I{}_J
		\end{pmatrix}\;,
	\end{equation} and therefore 
	\begin{equation}\label{Homogeneous space: complex compactified Minkowski}
		\CM \simeq \frac{\text{SL}(4\vert \mathcal{N})}{\text{P}_0}\;.
	\end{equation}
\end{itemize}

\subsubsection{Real case}

If we now look for the subgroups of \eqref{eqn: chiral poincare group 1}, \eqref{eqn: chiral poincare group 2} or \eqref{eqn: poincare group} that preserve the hermitian form
\begin{equation}\label{eqn: metric} h_{\overline{\hat{\alpha}}{\hat{\beta}}} = 
	\begin{pmatrix}
		0 & \mathbb{I}_2 & 0 \\
		\mathbb{I}_2 & 0 & 0 \\
		0 & 0 & -\delta^I{}_J
	\end{pmatrix}\;,
\end{equation}
i.e., the subset of matrices $P^{\hat\alpha}{}_{\hat\beta}$ such that $\overline{P}^{\overline{\hat\gamma}}{}_{\overline{\hat\beta}}{}\,h_{\overline{\hat\gamma}{\hat\delta}}\,P^{\hat\delta}{}_{\hat\alpha} = h_{\overline{\hat{\beta}}\hat{\alpha}}\,$, 
one finds a unique subgroup given by the set of matrices of the form 
\begin{equation}
	  	\text{P}_0^\mathbb{R} = \begin{pmatrix}
		{M^A}_B & 0^{AB'} & {0^A}_J \\
        i{{\overline{M}}_{A'}}{}^{C'} k_{C'B} + \frac{1}{2}{{S}}_{A'J}{} {(S^\dag)}^J{}_{C}{} M^C{}_B & -{{\overline{M}}_{A'}}{}^{B'} & S_{A'J} \\
        R^I{}_J{(S^\dag)}^J{}_A{} M^A{}_B & {0}^{IB'} & R^I{}_J
		\end{pmatrix}\;,
\end{equation}
where $R^I{}_J$ is unitary and $k_{C'B}$ is an arbitrary hermitian matrix.  
Note that if $M = (M^A{}_B)$, we have 
${M^\dag}_{A'}{}^{B'} = {\overline{M}}{}^{B'}{}_{A'}$. Therefore we recover the result that chiral and non-chiral superspaces are isomorphic in the real case, and their expression as homogeneous spaces for the real superconformal group is given by 
\begin{equation}\label{Homogeneous space: real compactified Minkowski}
	\CML^{\mathbb{R}} \simeq \CMR^{\mathbb{R}} \simeq {\CM}^{\mathbb{R}} \simeq \frac{ \text{SU}(2,2 \vert \mathcal{N})}{{\text{P}}_0^\mathbb{R}}\;.
\end{equation}

At the level of algebras, if we now look for the subset of $\mathfrak{su}(2,2\vert \mathcal{N})$ stabilising \eqref{eqn: origin twistor}, we get
\begin{equation}
    \mathfrak{p}_0^\mathbb{R} = \begin{pmatrix}
        m^A{}_B & 0^{AB'} & 0^A{}_J \\
        -i k_{A'B}  &-{((m)^\dagger)}_{A'}{}^{B'}& \overline{s}_{A'J} \\ 
        s^I{}_{B} & 0^{IB'}  & r^I{}_J
    \end{pmatrix} + c\,A\;,
\end{equation}
which is a parabolic subalgebra of $\su(2,2|\cN)\,$.

We say that a point $(Z^{\hat\alpha a}, \widetilde Z_{\hat\alpha}{}^b )$ in $\CM$ is real if the twistors are dual to each others 
through the metric \eqref{eqn: metric}, i.e.,
\begin{equation}
    \widetilde Z_{\hat\beta}{}^b =  {\overline Z}{}^{\bar{\hat\alpha} b} 
    h_{\bar{\hat\alpha} {\hat\beta}}\;. 
\end{equation}
The hermitian structure \eqref{eqn: metric} is invertible and we define the inverse $h^{\hat\alpha \bar{\hat\beta}}$ through 
$h_{\bar{\hat\alpha}\hat\alpha}\,h^{\hat\alpha \bar{\hat\gamma}} = 
 \delta_{\bar{\hat\alpha}}{}^{\bar{\hat\gamma}}\,$ or equivalently as
\begin{align}\label{eqn: real right}
    \widetilde{Z}_{\hat\alpha}\,h^{\hat\alpha \bar{\hat\beta}}\,
    \widetilde{Y}_{\bar{\hat\beta}}:= 
    \overline{Z}^{\bar{\hat\alpha}}\,h_{\bar{\hat\alpha}\hat\beta}\,Y^{\hat\beta}\;,
\end{align}
where $\widetilde{Y}_{\bar{\hat\alpha}} := h_{\bar{\hat\alpha} \hat{\beta}}\,
    Y^{\hat{\beta}}\; $.

Points of $\CML$ and $\CMR$ are  
real if the corresponding vector spaces are totally null with respect to the metric:
\begin{align}
\label{eqn:real cond super}
	\overline{Z}{}^{\bar{\hat\alpha} {a}}h_{\bar{\hat\alpha} 
 {\hat\beta}}Z^{\hat\beta b} 
	& = 0\,, \qquad \forall a,b\;,
 \\
\widetilde{Z}_{\hat{\alpha}}{}^b\,h^{\alpha\bar{\hat{\beta}}}\,
\overline{\widetilde{Z}}_{\bar{\hat{\beta}}}{}^a& =0 \;, \qquad \forall a,b\;.
\end{align} 
With these definitions, points in \eqref{Homogeneous space: real compactified Minkowski} are real.

\section{Invariant subspaces for the super Poincaré group}

In this section we break superconformal invariance and discuss 
the points left at infinity of Minkowski superspace by the action 
of the Poincar\'e supergroup.

\subsection{Super Poincaré group}

We will now break superconformal invariance down to the super Poincaré group by introducing infinity bi-supertwistors
\begin{align}\label{eqn: infinity bitwistors}
 I^{\hat{\alpha}\hat{\beta}} &= \begin{pmatrix}
     \epsilon^{AB} & 0 & 0 \\ 0& 0&0 \\ 0&0&0
 \end{pmatrix}, &  \tilde{I}_{\hat{\alpha}\hat{\beta}} &= \begin{pmatrix}
     0 & 0 & 0 \\ 0& \epsilon^{A'B'}&0 \\ 0&0&0
 \end{pmatrix}.
\end{align}
These bi-supertwistors are simple, $I^{\hat{\alpha}\hat{\beta}} = I^{[\hat{\alpha}}_1 I^{\hat{\beta}]}_2$,\  $I_{\hat{\alpha}\hat{\beta}} = I_{[\hat{\alpha}}^1 I_{\hat{\beta}]}^2$, and together define a point $(I^{\hat{\alpha}b}, \tilde{I}_{\hat{\alpha}}{}^b ) \in \CM$ given by
    \begin{align}
 \label{eqn: super null infinity twistor}
	I^{\hat{\alpha}b} &= \begin{bmatrix}
		\delta^{Ab} \\
		{0_{A'}}^b \\
		0^{Ib}
	\end{bmatrix}, &
	\tilde{I}_{\hat{\alpha}}{}^b &= \begin{bmatrix}
		 {0_{A}}^b &
		\delta^{A'b}&
		0^{Ib}
	\end{bmatrix}.
 \end{align}

The subgroup of \eqref{eqn: SL4} stabilising \eqref{eqn: infinity bitwistors} is parametrised by
\begin{equation}
    \begin{pmatrix}
        {M^B}_A & iT^{BA'} & {Q^B}_I \\
        0 & -{\widetilde{M}}_{B'}{}^{A'} & 0 \\
        0 & {\widetilde{Q}}^{JA'} & R^J{}_I
        \label{superpoinca}
    \end{pmatrix}  \;,
\end{equation}
where now ${M^B}_A,\ {{\widetilde{M}}_{B'}}{}^{A'} \in \text{SL}(2, \IC_c)$ and  $R^J{}_I \in \text{SL}( \cN, \IC_c)$. This is what we will call 
the complex super Poincar\'e group ${(\text{ISO}(1,3 \vert \mathcal{N}))}_{\mathbb{C}}\,$, 
where by an abuse of terminology we include the $R$-symmetry group 
$\text{SL}( \cN, \IC_c)\,$ in it.
The complex super Poincar\'e group has the structure 
$(\text{SL}(2,\IC)\times \text{SL}(2,\IC) \times \text{SL}(\mathcal{N})) \ltimes \mathcal{T}^{4\vert 4\mathcal{N}}$, where $\mathcal{T}^{4\vert 4\mathcal{N}}$ 
is the non-Abelian group consisting of the matrices
\begin{align}
    \begin{pmatrix}
    \delta^B{}_A & iT^{BA'} & Q^B{}_I \\
        0 & \delta_{B'}{}^{A'} & 0 \\
        0 & \widetilde{Q}^{JA'} & \delta^J{}_I
    \end{pmatrix}.
\end{align}

If we now impose that the group of matrices \eqref{superpoinca} should preserve 
the hermitian form \eqref{Hermitian form contraction}, i.e., 
that $P^{\hat\alpha}{}_{\hat\beta}$ should satisfy $\overline{P}^{\overline{\hat\gamma}}{}_{\overline{\hat\beta}}{}\,h_{\overline{\hat\gamma}{\hat\delta}}\,P^{\hat\delta}{}_{\hat\alpha} = h_{\overline{\hat{\beta}}\hat{\alpha}}$, 
we find a subgroup given by the set of matrices of the form 
\begin{equation}\label{eqn: real poincare}
	  	\text{P}^\mathbb{R} = \begin{pmatrix}
		{M^B}_A & iM^B{}_C t^{CA'}+\frac{1}{2}Q^B{}_I \overline{Q}^{IC'} \overline{M}_{C'}{}^{A'} & {Q^B}_I \\
        0_{B'A} & -\overline{M}_{B'}{}^{A'} & 0_{B'I} \\
        0^J{}_A & {-R^J{}_I\overline{Q}^{IB'}}{{\overline{M}}_{B'}}{}^{A'} & R^J{}_I
		\end{pmatrix}\;,
\end{equation}
where $R^I{}_J$ is unitary and $t_{CA'}$ is hermitian. 
We call this subgroup the real super Poincaré group.

\subsection{Minkowski superspace}

A point of the superconformal compactification is 
at infinity whenever $\det( \tilde I_{\hat \alpha}{}_a Z^{\hat\alpha b} = {\pi_{a}}^{b}) \neq 0$, 
on the contrary points of Minkowski superspace are those satisfying  
\begin{equation}
\det( \tilde I_{\hat \alpha}{}_a Z^{\hat\alpha b} = {\pi_{a}}^{b}) \neq 0\;.
\end{equation}

\subsubsection{Chiral superspaces}

The chiral left Minkowski superspace ${M}_\L$ is defined \cite{manin_gauge_1988} as the subset of points $ Z^{\hat\alpha b} \in \CML$ 
such that $\det( \tilde I_{\hat \alpha}{}_a Z^{\hat\alpha b} = {\pi_{a}}^{b}) \neq 0$. In this chart, 
the coordinates are such that ${\pi_{A'}}^b$ can be set to the identity, and we write
\begin{equation}\label{eqn: twist mink}
  Z^{\hat\alpha b} = \begin{bmatrix}
		iX^{Ab}_+ \\
		{\delta_{A'}}^b\\
		\theta^{Ib}
	\end{bmatrix}.
\end{equation}
The second line allows to identify $A'$ with $b$, and $( X^{AA'}_+ , \theta^{IA'} )$ 
form the chiral coordinates on $M_\L\,$.
Similarly, we have chiral coordinates $( X^{A'A}_-, \tilde\theta^{IA} )$ on $M_\R\,$:
\begin{equation}\label{dual twist mink}
  \widetilde{Z}_{\hat\alpha}{}^b = \begin{bmatrix}
		{\delta_{A}}^b &
  -iX^{A'b}_- &
		-\tilde\theta^{b}_I
	\end{bmatrix}.
\end{equation}

The action of the super Poincaré group \eqref{superpoinca} on $M_\L$ is given by 
\begin{equation}
\begin{split}
	\begin{pmatrix}
        {M^B}_A & iT^{BA'} & {Q^B}_I \\
        0 & -{{\widetilde{M}}_{B'}}{}^{A'} & 0 \\
        0 & {\widetilde{Q}}^{JA'} & R^J{}_I
    \end{pmatrix} 
    \begin{bmatrix}
		iX^{Ab}_+ \\
		{\delta_{A'}}^b\\
		\theta^{Ib}
	\end{bmatrix} &= \begin{bmatrix}
		{M^B}_A iX^{Ab}_+ + iT^{Bb} + Q^B{}_I\theta^{Ib} \\
		-\widetilde{M}_{B'}{}^b \\
		\widetilde{Q}^{Jb} + R^J{}_I\theta^{Ib}
	\end{bmatrix}\\
	& \sim
	\begin{bmatrix}
		{(\widetilde M)}^{b}{}_{c}\cdot ({M^B}_A iX_+^{Ac}+ iT^{Bc} + Q^{B}{}_I\theta^{Ic})\\
		{\delta_{B'}}^b \\
		({\widetilde{M}})^b{}_c(\widetilde{Q}^{Jc} + R^J{}_I\theta^{Ic})
	\end{bmatrix} \in M_\L\;.
\end{split}
\end{equation}
The action is transitive which makes ${M}_\L$ an homogeneous space for the complex super Poincaré group:
\begin{equation}\label{homogeneous space: complexified minkowski}
	M_\L \simeq \frac{{(\text{ISO}(1,3 \vert \mathcal{N}))}_{\mathbb{C}}}{(\text{SL}(2,\mathbb{C}_c)\times \text{SL}(2,\mathbb{C}_c) \times \text{SL}(\cN, \mathbb{C}_c)) \ltimes \IC^{0\vert 2 \cN}}\;.
\end{equation}
The stabiliser is obtained from \eqref{superpoinca} by 
fixing the point  $X_+^{AA'} = 0$, $\theta^{Ib} = 0$ of ${M}_\L$ and 
this imposes $T^{BA'} = 0$, $\widetilde{Q}^{JA'} = 0$. 

For $M_\R$ it is obtained in a similar way by asking to fix the point  
$X_-^{A'A} = 0\,$, $\theta^{Ib} = 0\,$ and this imposes instead $T^{BA'} = 0\,$, 
${Q}^{B}{}_I = 0\,$. Therefore, $M_\L$ and $M_\R$ have the same expression 
\eqref{homogeneous space: complexified minkowski} as homogeneous spaces, 
although they are not the same space.

\paragraph{Reality conditions.}
If we restrict ourselves to the real case, i.e., 
if we impose on chiral left Minkowski superspace the condition 
\eqref{eqn:real cond super}, 
we get 
\begin{align}
	X^{AA'}_+ - \overline{X}{}^{A'A}_+ 
 = -i \,\overline{\theta}^{IA}\,\delta_{IJ}\,\theta^{JA'}
 \,.
\end{align}
As a result of this reality condition, we conclude that there 
exists a hermitian matrix $x^{AA'}$ such that
\begin{equation}
	X_+^{AA'} = x_\L^{AA'} - \frac{i}{2}\, 
 \overline{\theta}^{IA}\,\delta_{IJ}\,\theta^{JA'}.
 \label{X+}
\end{equation}
Similarly, when we impose the reality conditions 
$\overline{Z}_{\bar{\hat{\alpha}}}{}^a \,h^{\bar{\hat{\alpha}}\beta} \,Z_{\hat{\beta}}{}^b
=0$
on chiral right Minkowski superspace, consistently with \eqref{eqn: real right},
we find that there exists a hermitian matrix $x_r^{AA'}$ such that
\begin{equation}
	X_-^{A'A} = x_\R^{AA'} + \frac{i}{2} \,\overline{\widetilde\theta}{}^{\,IA}\delta_{IJ}{\widetilde\theta}^{JA'}\;.
 \label{X-}
\end{equation}
The real Poincaré group \eqref{eqn: real poincare} acts transitively 
on these real Minkowski superspaces. 
Accordingly, \eqref{homogeneous space: complexified minkowski} reduces to
\begin{equation}\label{homogeneous space: real minkowski}
	{M}_{\mathbb{R}}^{4\vert 4\mathcal{N}} \simeq \frac{\text{ISO}(1,3 \vert \mathcal{N})}{\text{SL}(2,\IC_c) \times \text{SU}(\cal N)}\;.
\end{equation}

\subsubsection{Non-chiral superspace}
 The non-chiral Minkowski superspace $M$ is defined via the chiral left and right Minkowski superspaces :
 \begin{equation}
 	M = \{ (Z^{\hat{\alpha}b}, {\widetilde{Z}}_{\hat{\alpha}b}) \mid Z^{\hat{\alpha}b} \in M_\L \text{ and } {\widetilde{Z}}_{\hat{\alpha}b} \in M_\R \text{ such that } \widetilde{Z}_{\hat{\alpha}}{}^b Z^{\hat{\alpha}a} = 0 \}\;.
 \end{equation}
The flag condition between \eqref{eqn: twist mink} and \eqref{dual twist mink} imposes the relation 
\begin{equation}\label{eqn: flag cond}
	X_+^{AA'} - X_-^{A'A} = -i\tilde{\theta}_{I}^{A}{\theta}^{A'I}
\end{equation}
or, equivalently,
\begin{align}
    X_+^{AA'} &= x^{AA'} - \frac{i}{2} \tilde{\theta}_{I}^{A}{\theta}^{A'I}, & 
    X_-^{A'A} &= x^{AA'} + \frac{i}{2} \tilde{\theta}_{I}^{A}{\theta}^{A'I}\;,
\end{align}
where $(x^{AA'},{\theta}^{A'I}, \tilde{\theta}_{I}^{A})$ are coordinates on $M\,$. 
The group \eqref{superpoinca} acts on $M$ and the stabiliser is the intersection of the stabilisers of the chiral spaces $M_\L$ and $M_\R$. 
One recovers that complex Minkowski superspace is an homogeneous space for the complexified super Poincaré group.

\begin{equation}
	M \simeq \frac{{(\text{ISO}(1,3 \vert \mathcal{N}))}_{\mathbb{C}}}{\text{SL}(2,\mathbb{C}_c)\times \text{SL}(2,\mathbb{C}_c) \times \text{SL}(\cN, \mathbb{C}_c) }\;.
\end{equation}

\paragraph{Reality conditions.}
The real points are obtained
by imposing on the points $$\{(Z^{\hat\alpha b}, \widetilde{Z}_{\hat\alpha}{}^b)
~|~ {\widetilde{Z}}_{\hat{\alpha}}{}^b Z^{\hat{\alpha}a}= 0\}$$ 
of complex non-chiral Minkowski space the reality conditions 
$\widetilde{Z}_{\hat\alpha}{}^b = 
h_{\overline{\hat\alpha}\hat\alpha}\overline{Z}^{\overline{\hat\alpha} b} $, 
where the complex structure $h_{\overline{\hat\alpha} \beta}$ 
is given in \eqref{eqn: metric}.
This is equivalent to $X_-^{A'A} = \overline{X}_+{}^{A'A}\,$ 
(i.e., $x^{AA'} = \overline{x}^{A'A}$) and
$\tilde{\theta}^{IA} = \overline{\theta}{}^{IA}\,$.
The corresponding real homogeneous space is again \eqref{homogeneous space: real minkowski}.

\subsubsection{Summary of the coordinates}

Keeping in mind that  $X_+^{AA'} = x_\L^{AA'} - \frac{i}{2}\, 
 \overline{\theta}^{IA}\,\delta_{IJ}\,\theta^{JA'}$
 and $X_-^{A'A} = x_\R^{AA'} + \frac{i}{2} \,\overline{\widetilde\theta}{}^{\,IA}\delta_{IJ}{\widetilde\theta}^{JA'}\,$ the situation can be summarised by the following.

\vspace{0.1cm}
{\renewcommand{\arraystretch}{1.5}
\begin{tabular}{c|c|c}
	\ & Complex & Real \\[0.1em] \hline  
	Chiral left  & $(X_+^{AA'}, \theta^{IA})$ & $(x_\L^{AA'},\theta^{IA})$ with $x_\L^{AA'}$ hermitian  \\ 
	(resp. right) & (resp. $(X_-^{AA'}, \tilde\theta^{IA'})$) & (resp. $(x_\R^{AA'},\tilde\theta^{IA'})$) with $x_\R^{AA'}$ hermitian \\[0.2em] \hline
 Non-chiral & $(x^{AA'}, \theta^{IA'}, \tilde\theta^A_I)$ & $(x^{AA'}, \theta^{IA'})$ (with $\tilde\theta = \bar\theta$ and $x_\L^{AA'} = x_\R^{AA'} = x^{AA'}$ hermitian) 
\end{tabular}}

\subsection{Super Null Infinity}

We will now discuss points at infinity, i.e., satisfying 
$\det({\pi_{A'}}^{b}) = 0$. Among those, points for which ${\pi_{A'}}^{b} = 0$ correspond to a supersymmetric version of space/time like infinity which we will here exclude. Under this assumption can always make use of 
 $\text{GL}(2,\IC)$ invariance in the choice of the generators of plane to write
\begin{equation}\label{Points of super scri}
	Z^{\hat{\alpha} b} = \begin{bmatrix}
		\omega^A & \hat{\pi}^A \\
		\pi_{A'} & 0 \\
		\theta^I & \eta^I=0 
	\end{bmatrix}.
\end{equation}
The condition $\eta^I =0$ in the above equation is a Poincaré invariant condition which is part of the definition of chiral left super null infinity $\mathscr{I}_\L$ (using dual twistors, chiral right super null infinity $\mathscr{I}_\R$ can be defined exactly in the same way). The most general situation with $\eta^I \neq0$ correspond to another invariant subspace which will be investigated in the next section.

\subsubsection{Chiral superspaces}

Using the remaining $\text{GL}(2,\IC)$ freedom, we can pose
\begin{equation}
	u_+ = -i\omega^A \hat{\pi}_A\;,
\end{equation}
which can be interpreted as a chiral left coordinate on $\mathscr{I}_\L$, on which now there are coordinates $(u_+, \hat{\pi}^A, \pi_{A'}, \theta^I)\in \IC \times \mathbb{CP}^1 \times \mathbb{CP}^1 \times \IC^{0\vert \cN}$. Similarly, on $\mathscr{I}_\R$, we have coordinates 
	$(u_-, \tilde{\hat\pi}^A, \tilde\pi_{A'}, \tilde\theta^I)$.

The action of the super Poincaré group \eqref{superpoinca} on $\mathscr{I}_\L$ is given by 
\begin{equation}
	\begin{pmatrix}
		{M^B}_A & iT^{BA'} & {Q^B}_I \\
        0 & -{\widetilde{M}}_{B'}{}^{A'} & 0 \\
        0 & {\widetilde{Q}}^{JA'} & R^J{}_I
	\end{pmatrix}
	\cdot
	\begin{bmatrix}
		\omega^A & \hat{\pi}^A \\
		\pi_{A'} & 0 \\
		\theta^I & 0^I
	\end{bmatrix} = 
	\begin{bmatrix}
		{M^B}_A \omega^A + iT^{BA'}\pi_{A'} + Q^B{}_I\theta^I &  {M^B}_A\hat{\pi}^A\\
	-\widetilde{M}_{B'}{}^{A'} \pi_{A'} & 0 \\
	\widetilde{Q}^{JA'}\pi_{A'} +  R^J{}_I \theta^I & 0^I
	\end{bmatrix} \in \mathscr{I}_\L\;.
\end{equation}
On the chiral right $\mathscr{I}_\R$, it acts by the inverse as described in \eqref{eqn: action}. 

The action is transitive which makes ${\mathscr{I}}_\L$, ${\mathscr{I}}_\R$ both homogeneous spaces for the complex super Poincaré group. 
The stabiliser, which is constituted of matrices of the form
\begin{equation}
  \begin{pmatrix}
  	M^1{}_1 & M^{1}{}_{2} & T^{11} & T^{12} & Q^1{}_I  \\
  	0 & (M^1{}_1)^{-1} & T^{21} & 0 & Q^2{}_I \\
  	0 & 0 & -\widetilde{M}_1{}^1 & 0 & 0_I\\
  	0 & 0 & \widetilde{M}_2{}^1 & -(\widetilde{M}_1{}^1)^{-1} & 0_I\\
  	0^I & 0^I & \tilde{Q}^{1I} & 0^I & R^J{}_I
  \end{pmatrix}\;,
 \end{equation}
fixes $M^2{}_2$, $M^2{}_1$, $\widetilde{M}_2{}^2$, $\widetilde{M}_1{}^2$, $T^{22}$ and  $\widetilde{Q}^{2I} = 0$
(as compared to a linear combination of $Q^{1I}$ and $Q^{2I}$ for the right space) 
and parametrises in both cases the subgroup $\IC^3\rtimes(\mathbb{C}^{0\vert 3\mathcal{N}}\rtimes(\text{ISO}(2) \times \mathbb{R}^* \times \text{ISO}(2) \times \mathbb{R}^* \times \text{SL}(\mathcal{N})))$.
This can be shown by asking to stabilise the point 
\begin{equation}
	\begin{bmatrix}
  	0 & 1 \\
  	0 & 0 \\
  	0 & 0 \\
  	1 & 0 \\
  	0^I & 0^I
  \end{bmatrix} \in \mathscr{I}_\L\;,
\end{equation}
(respectively the point
$\begin{bmatrix}
	0 & 1 & 0&0 & 0^I \\
	0 & 0 & 0&1 & 0^I
\end{bmatrix} \in \mathscr{I}_\R$).

Therefore, we obtain the expression 
\begin{equation}\label{Homogenous space: complexified left superscri}
  \mathscr{I}_\L \simeq \mathscr{I}_\R \simeq \frac{{(\text{ISO}(1,3 \vert \mathcal{N}))}_{\mathbb{C}}}{\IC^{3|0}\rtimes(\mathbb{C}^{0\vert 3\mathcal{N}} \rtimes(\text{ISO}(2) \times \mathbb{R}^* \times \text{ISO}(2) \times \mathbb{R}^* \times \text{SL}(\mathcal{N})))}\;.
\end{equation}

\paragraph{Reality conditions.}
By imposing \eqref{eqn:real cond super}, we get the 
following two conditions
\begin{align}\label{eqn: reality cond on scrai}
	[\hat{\pi}_A] &= [\overline{\pi}_A] \;, &
	-2\,\text{Im}(u_+) - \overline\theta\theta &= 0\;, 
\end{align}
and so we have local coordinates 
$(u_\L, ([\pi^{A'}],[\bar\pi^{A}]) , \theta^I) \in \IR \times S^2 \times \IC^{0\vert \cN}$, where $u_+ = u_\L - \frac{i}{2}\overline\theta^I\theta_I$  (similarly on $\mathscr{I}_\R$ we have 
$(u_\R, ([\tilde\pi^A], [\overline{\tilde\pi}^{A'}]), \tilde\theta^I)$ 
with $u_- = u_\R + \frac{i}{2}\overline{\tilde\theta}^I\tilde\theta_I$).

By intersecting with the subgroup \eqref{eqn: real poincare} that acts transitively on the real points of $\mathscr{I}$, we obtain the expression for the real super null infinity
  \begin{equation}\label{Homogeneous space: real super null infinity}
  	  \mathscr{I}_\IR^{3\vert 2\mathcal{N}} \simeq \frac{{\text{ISO}(1,3 \vert \mathcal{N})}}{\IR^3\rtimes(\mathbb{R}^{0\vert 2\mathcal{N}}\rtimes(\text{ISO}(2) \times \mathbb{R}^* \times \text{SU}(\mathcal{N})))}\;.
  \end{equation}
  
\subsubsection{Non-chiral superspace}

We define the non-chiral super null infinity as the space $\mathscr{I}_\L \times \mathscr{I}_\R \in F(2\vert 0, 2\vert \mathcal{N}, \IC^{4\vert \mathcal{N}})$.
In coordinates, the flag conditions on 
\begin{align}
    (Z^{\hat\alpha a }, \tilde Z^a{}_{\hat\alpha}) = 
    \left(
    \begin{bmatrix}
        \omega^A & \hat\pi^A \\
        \pi_{A'} & 0_{A'} \\
        \theta^I & 0^I
    \end{bmatrix}\;,
    \begin{bmatrix}
         \tilde\pi_A & \tilde\omega^{A'} & -\tilde\theta_I \\
        0_A & \tilde{\hat\pi}^{A'} & 0_I
    \end{bmatrix}  \right) \in \mathscr{I}_\L \times \mathscr{I}_\R
\end{align} 
are realised by 
\begin{align}
    \tilde{\hat\pi}^{A'}\,\pi_{A'}  &= 0\;,
    \label{flag I nonchiral1}\\
    \tilde{\pi}_A\,\omega^A + \tilde{\omega}^{A'}\,\pi_{A'}-\tilde\theta^I\,\theta_I &=0\;,\label{flag I nonchiral2}\\
    \tilde{\pi}_A\,\hat{\pi}^A  &= 0\;.
    \label{flag I nonchiral3}
\end{align}

Let us introduce again the GL$(2,\mathbb{C})$ invariants 
\begin{align}
 iu_+ &:= \omega^A\,\hat{\pi}_A\;, &
 -iu_- &:= \tilde\omega^{A'}\,\tilde{\hat{\pi}}_{A'}\;.
\end{align}

Then the flag conditions \eqref{flag I nonchiral2} -- \eqref{flag I nonchiral3} are solved by 
\begin{align}\label{eqn: scrai flag}
	[\tilde{\pi}_A] &= [\hat{\pi}_A] & [\tilde{\hat\pi}_{A'}] &= [{\pi}_{A'}] & iu_+ - iu_- -\tilde\theta^I \theta_I= 0\;.
\end{align}

Setting 
\begin{align}
    u_+ &= u_l - \frac{i}{2}\,\tilde\theta^I \theta_I\;, &
    u_- &= u_r + \frac{i}{2}\,\tilde\theta^I \theta_I\;,
\end{align}
one solves the condition \eqref{eqn: scrai flag} with $u_l = u_r=u\,$.
Non-chiral coordinates on complex $\mathscr{I}_\L \times \mathscr{I}_\R$ are then given by 
$(u, [\pi_{A'}], [\tilde{\pi}_A],  \theta^I, \tilde\theta^I )$.
\paragraph{Reality conditions.}

The real points are obtained
by imposing on the points $\{(Z^{\hat\alpha b}, \widetilde{Z}_{\hat\alpha}{}^b)
~|~ {\widetilde{Z}}_{\hat{\alpha}}{}^b Z^{\hat{\alpha}a}= 0\}$ 
of complex non-chiral $\mathscr{I}_\L \times \mathscr{I}_\R$ the reality conditions 
$\widetilde{Z}_{\hat\alpha}{}^b = 
h_{\overline{\hat\alpha}\hat\alpha}\overline{Z}^{\overline{\hat\alpha} b} $, 
where the complex structure $h_{\overline{\hat\alpha} \beta}$ 
is given in \eqref{eqn: metric}.
This is equivalent to $\tilde{\pi}_A = \overline{\pi}_A$,
$\tilde{\theta}^{I} = \overline{\theta}{}^{I}\,$
and $$\overline{iu_+} = \overline{\omega}^{A'}\overline{\hat\pi}_{A'} = \tilde{\omega}^{A'}\tilde{\hat\pi}_{A'} = iu_- \;,$$
therefore $\overline{u_+} = u_-$ and this implies $\overline{u} + \frac{i}{2}\overline{\theta}^I\theta_I = u + \frac{i}{2}\overline{\theta}^I\theta_I $, i.e. $u \in \IR$.

The corresponding real homogeneous space is again \eqref{Homogeneous space: real super null infinity}.

\subsubsection{Summary of the coordinates}

Keeping in mind the notation $u_+ = u_\L - \frac{i}{2}\overline\theta^I\theta_I$ and 
$u_- = u_\R + \frac{i}{2}\overline{\tilde\theta}^I\tilde\theta_I$, the coordinates on super null infinity are summarised as follows:

\vspace{0.1cm}
{\renewcommand{\arraystretch}{1.8}
\begin{tabular}{c|c|c}
	\ & Complex & Real \\[0.1em] \hline
	Chiral left  & $(u_+, [\pi_{A'}] , [\hat{\pi}^A], \theta^I)$ & $(u_\L, {[\pi_{A'}]}, \theta^I)$ with $u_\L \in \IR$
 and $\hat{\pi}^A=\bar\pi^A$
 \\ 
	(resp. right) & (resp. $(u_-, [\tilde\pi_{A}] , [\tilde{\hat{\pi}}^{A'}], \tilde\theta^I)$) & (resp. $(u_\R, [\tilde\pi_A],\tilde\theta^{I})$ with $u_\R \in \IR$ and $\tilde{\hat \pi}^{A'}=\bar {\tilde \pi}^{A'}$) \\[0.2em] \hline
	Non-chiral & $(u, [\pi_{A'}], [\tilde{\pi}_A],  \theta^I, \tilde\theta^I )$ & $(u, [\pi_{A'}], \theta^{I})$ with $u \in \IR$, $\tilde{\pi}^A=\bar\pi^A$ and $\tilde \theta^I = \bar \theta^I$ 
\end{tabular}}

\subsection{Other invariants subspaces}

\subsubsection{Subspace \texorpdfstring{$\cal{H}$}{}}
\label{subsubsec: H}

This subspace is defined in the chiral left sector as the set of points with  $\det({\pi_{A'}}^{b}) = 0$, ${\pi_{A'}}^{b} \neq 0$ but where, as compare to \eqref{Points of super scri}, $\eta$ is supposed to be non zero: 
\begin{equation}
	Z^{\hat{\alpha} b} = \begin{bmatrix}
		\omega^A & \hat{\pi}^A \\
		\pi_{A'} & 0 \\
		\theta^I & \eta^I\neq0 
	\end{bmatrix}.
\end{equation}

The $\text{GL}(2,\IC)$-invariant quantity
\begin{equation}
	i u_+ = \omega^A \hat{\pi}_A\;,
\end{equation}
can then be interpreted as one of the chiral left coordinates $(u_+, [\pi_{A'}], [\hat{\pi}^A], \theta^I, \eta^I)\in \IC \times \mathbb{CP}^1 \times \mathbb{CP}^1 \times \IC^{0\vert \mathcal{N}} \times \IC^{0\vert \mathcal{N}}$ on $\mathcal{H}_\L$.
This sub-supermanifold is invariant under super Poincar\'e. However the action is \emph{not} transitive because the $R$-symmetry $\mathrm{SL}(\cN,\IC_c)$ does not act transitively on $\IC^{\cN}_a$. Let us here discuss the orbits of the particular point 
\begin{align}
	Z^{\hat{\alpha} b} &= \begin{bmatrix}
  	0 & 1 \\
  	0 & 0 \\
  	0 & 0 \\
  	1 & 0 \\
  	0^I & \eta^I
  \end{bmatrix} & \text{with}\qquad  \eta^I &=\begin{pmatrix} a \\0 \\ \vdots\\ 0 \end{pmatrix}.
\end{align}
For $\cN=1$ all points of $\mathcal{H}_\L$ lie in such an orbit and 
$a\in \IC_a$ parametrises the different possible orbits; for $\cN\geq 2$ there 
are more complicated orbits of the type $\eta^I = ( a ,\,b ,\, \dots\,, 0)$ with 
$ab\neq 0$. As we shall see the situation for $\cN=1$ is already interesting enough. 
The action of the super Poincaré group is given by
\begin{equation}
\begin{bmatrix}
  	0 & 1 \\
  	0 & 0 \\
  	0 & 0 \\
  	1 & 0 \\
  	0 & a \\
  	0 & 0 \\
  	\vdots & \vdots \\
  	0 & 0
  \end{bmatrix} \mapsto 
  \begin{pmatrix}
  	M^{1}{}_1 & M^{1}{}_2 & iT^{11'} & iT^{12'} & Q^1{}_1 & \cdots & Q^1{}_N \\
  	M^{2}{}_1 & M^{2}{}_2 & iT^{21'} & iT^{22'} & Q^2{}_1 & \cdots & Q^2{}_N \\
  	0 & 0 & -\widetilde{M}_{1'}{}^{1'} & -\widetilde{M}_{1'}{}^{2'} & 0 & \cdots & 0 \\
  	0 & 0 & -\widetilde{M}_{2'}{}^{1'} & -\widetilde{M}_{2'}{}^{2'} & 0 & \cdots & 0\\
  	0 & 0 & \widetilde{Q}^{11'} & \widetilde{Q}^{12'} & R^1{}_1 & \cdots & R^1{}_N \\
  	\vdots & \vdots & \vdots & \vdots & \vdots & \ddots & \vdots \\
  	0 & 0 & \widetilde{Q}^{N1'} & \widetilde{Q}^{N2'} & R^N{}_1 & \cdots & R^N{}_N
  \end{pmatrix} \cdot
  \begin{bmatrix}
  	0 & 1 \\
  	0 & 0 \\
  	0 & 0 \\
  	1 & 0 \\
  	0 & a \\
  	0 & 0 \\
  	\vdots & \vdots \\
  	0 & 0
  \end{bmatrix}\;,
  \end{equation}
and the subgroup stabilising our preferred point requires the following equations to be satisfied :
\begin{align}
	iT^{22'} &= 0 \;,&& \\
	\widetilde{M}_{1'}{}^{2'} &= 0\;,&&\\
	\widetilde{Q}^{12'} &= i T^{12'} a\;,\\  
 \widetilde{Q}^{J2'} &= 0 \quad \qquad\text{ if } J>2\;,&& \\
	M^{2}{}_1& = -Q^2{}_1 a \;,& \\
M^1{}_1 &= R^1{}_1 + \mathfrak{M}^1 \,a \;,\\ 
R^J{}_1 &= \mathfrak{R}^J\,a \quad \qquad 	\text{ if } J>2\;, 
\end{align}
where $\mathfrak{M}^1$ and $\mathfrak{R}^J$ are some unconstrained anti-commuting numbers. Let us now stick to $\cN=1$ in order to simplify the discussion: The parameters left unconstrained for the stabiliser are then $M^1{}_2$, 
$\widetilde M^1{}_1$, $\widetilde M^1{}_2$, $T^{11'}$, $T^{12'}$, $T^{21'}$, $Q^{A}$ and $\mathfrak{M}^1$. One might be tempted to conclude that this stabiliser is a supergroup of dimension $\IC^{6|3}$ and therefore that the orbits an homogeneous space of dimension $\IC^{4|1}$. 
(It might here be useful for the reader to compare with $\mathscr{I}_l$ which is an homogeneous space \eqref{Homogenous space: complexified left superscri} of dimension $\IC^{3|\cN}$.)
Strictly speaking this would however be an incorrect conclusion: the fact that the ``coordinate" $\mathfrak{M}^1$ here only appears as the product ``$\mathfrak{M}^1 \,a$" means that it is ambiguous and that, following DeWitt \cite{dewitt_supermanifolds_1992}, it does not define a superchart; the stabiliser is 
in fact strictly speaking not a supermanifold. If the same analysis had been instead 
performed at the level of the algebra one would have equivalently concluded that the 
corresponding stabiliser is a sub supervector space 
(over $\Lambda_\infty$)
of the super Poincaré algebra \emph{which does not admit a basis}.

We here wish to comment on the peculiar features of supergeometries which are responsible for these phenomena.
If one follows DeWitt's approach to supermanifolds 
(this approach is equivalent to the algebro-geometric approach, see 
\cite{rogers2007supermanifolds} for a proof), then the supervector spaces on which 
manifolds are modelled are constructed over the supernumbers $\Lambda_\infty$, 
which form a graded ring and a graded infinite dimensional vector space over $\IC$.  
Nevertheless, because theses supervector spaces of DeWitt are not over a field 
but rather over the graded \emph{ring} $\Lambda_\infty$, they are by definition (graded) 
modules. 
In general, the analogy with vector space is dangerous because 
not all modules 
over an algebra $\mathcal{A}$ admit a basis.  In the 
case where a (graded) module over $\mathcal{A}$ admit a basis it is said to 
be free and the parallel with vector spaces is possible: in \cite{tuynman2004supermanifolds} this analogy 
between vector spaces and free modules (there called $\mathcal{A}\,$-vector spaces) has 
been investigated with great care. This, nevertheless, has certain important limits.
The main limit, that turns out to be crucial for our analysis, is the notion of 
subobject. 
In fact, there is no natural way to induce a basis on a submodule, 
once a basis for the free graded $\mathcal{A}\,$-module is given,  
due to the counter-intuitive fact that submodules of 
a free module do not have to be free.  
One therefore \cite{tuynman2004supermanifolds} restricts the notion of a graded 
subspace in this new category as being a 
graded subspace in the usual sense, with the additional restriction 
that there exists a homogeneous basis for the total space within 
its equivalence class such that a subset of it 
forms a basis for the graded subspace in question; 
these are the sub-$\mathcal{A}$-vector spaces of \cite{tuynman2004supermanifolds}. 
This being introduced, supermanifolds and sub-supermanifold are respectively 
modelled on $\Lambda_{\infty}$-vector spaces and sub-$\Lambda_{\infty}$-vector spaces.
However, one should realise that, contrary to the ordinary case, one cannot 
guarantee that through every point of a supermanifold passes a sub-supermanifold 
of lower dimension \cite{tuynman2004supermanifolds}.

Note however that, despite the fact that these orbits are not a supermanifold, this is very easy to describe them explicitly as 
\begin{equation}
	Z^{\hat{\alpha} b} = \begin{bmatrix}
		\omega^A & \hat{\pi}^A \\
		\pi_{A'} & 0 \\
		\theta^I & \alpha a 
	\end{bmatrix}\;.
\end{equation}
Here $a \in \IC_a$ is fixed and parametrises different orbits while 
$\alpha \in \IC_c$ is an extra commuting ``coordinate" parametrising the 
would-be extra bosonic dimension of these orbits as compared to 
$\mathscr{I}_{\ell}$.

As we shall see reality conditions complicate further the situation since 
neither the real orbits nor the real subspace of $\mathcal{H}$ are 
sub-supermanifolds. The non-chiral models have the same problem.

\paragraph{Reality conditions.} 
We are going to impose the following reality conditions:
\begin{subequations}
\begin{align}
	{(Z^1)}^2 = 0 & \Rightarrow \omega^A{\overline{\pi}}_A 
	+ \pi_{A'}{\overline{\omega}}^{A'}+\delta_{IJ}\theta^{I}\overline{\theta}^{{J}} = 0 
	\textit{ i.e. } 2\,\text{Re}(\omega^A\overline{\pi}_A) 
	- \delta_{IJ}\overline{\theta}^{{J}}\theta^{I} = 0\;, \label{H: first reality condition}\\
	{(Z^2)}^2 = 0 & \Rightarrow \delta_{IJ}\eta^{I}\overline{\eta}^{J} = 0\;, \label{H: second reality condition}\\
	{Z^1 \cdot Z^2} = 0 & \Rightarrow \overline{\pi}_A \hat{\pi}^A \label{H: third reality condition}
	-\overline{\theta}_{J}\eta^{J} = 0 \;.
\end{align}
\end{subequations}
The same type of technical problem we met in the context of 
the stabiliser of $\cal H$ shows up at this stage: 
The second equation above does not define a sub-supermanifold 
and therefore, here again, strictly speaking the real 
supermanifold $\mathcal{H}$ cannot be defined.

For $\mathcal{N} = 1\,$, 
we can  solve the reality conditions and deduce a parametrisation of the real $\mathcal{H}\,$: First, equation \eqref{H: second reality condition} becomes $\eta \bar \eta =0$ and is solved as
$\eta= e^{i\phi}\,a$, 
with $a\in \mathbb{R}_a$ a real anticommuting supernumber and $\phi \in\mathbb{R}_c$ a 
real commuting supernumber. Then, in order to solve equation \eqref{H: third reality condition}, we introduce the parametrisation $\eta = \eta_{A}\hat{\pi}^{A}$, 
where the odd variables $\eta_{A}$ is chosen to satisfy 
$\eta_{A}\bar{\eta}_{A'}=0$ (this ensures ensure that 
\eqref{H: second reality condition} still holds). Such a decomposition 
of $\eta$ is not unique but always exists. 
In fact, because $\hat{\pi}^A$ has non zero body, at least one of the body of the 
components 
in $\{\hat{\pi}^0, \hat{\pi}^1\}$ is non zero. 
Assume for example that $\hat{\pi}^0$ has non zero body;
then we can take $\eta_A = (\eta/\hat{\pi}^0, 0)$. With this choice of 
parametrisation, the third reality condition \eqref{H: third reality condition} 
reads
\begin{equation}
    ( \bar{\pi}_A -\bar{\theta}\eta_{A} ) \hat\pi^A=0\;,
\end{equation}
and, since $\hat{\pi}^A$ has a non vanishing body, can be solved as
\begin{equation}
    \hat\pi^A = \alpha (\bar{\pi}^A - \bar{\theta}\eta^{A}) \;,
    \qquad \alpha \in \mathbb{C}\;.
\end{equation}
Using complex projective invariance to rescale $Z^{\hat{\alpha}2}$, 
one gets
\begin{equation}
	Z^{\hat{\alpha} b} = \begin{bmatrix}
		\omega^A & \bar{\pi}^A - \bar{\theta}\eta^{A}  \\
		\pi_{A'} & 0 \\
		\theta & 
      \eta{}_A(\bar{\pi}^A - \bar{\theta}\eta^{A})
	\end{bmatrix}.
\end{equation}
Finally, in order to solve the first reality condition \eqref{H: first reality condition} we introduce $\Theta := \theta + \eta^A\omega_A$ and $iu_+ := \omega^{A}(\bar{\pi}_A
	- \bar{\Theta}\eta_{A} )$. We then have
\begin{equation}
	Z^{\hat{\alpha} b} = 
\begin{bmatrix}
		\omega^A & \bar{\pi}^A - \bar{\Theta}\eta^{A} \\
		\pi_{A'} & 0 \\
		\Theta + \omega^A\eta_A & 
     \eta_A(\bar{\pi}^A-\bar{\Theta}\eta^{A})
	\end{bmatrix} = 
 \begin{bmatrix}
		\omega^A & \bar{\pi}^A - \bar{\Theta}\eta^{A} \\
		\pi_{A'} & 0 \\
		\Theta + \omega^A\eta_A & e^{i\phi} a
	\end{bmatrix}\;,
\end{equation}
and equation \eqref{H: first reality condition} becomes
\begin{equation}
    2 \,Im(u_+) = \Theta \bar{\Theta}\;.
\end{equation}
Which is solved as
\begin{equation}
    u_+ =  u_\L + \frac{i}{2} \Theta \bar{\Theta}\;,\quad u_\L \,\in\,\mathbb{R}_c\;.
\end{equation}
We thus found a parametrisation of $\mathcal{H}\,$ by
$(u_\L,[\pi_{A'}], e^{i\phi}, \Theta, a ) \in \IR \times \mathbb{CP}^1 \times S^1  \times \IC_a \times \IR_a$. 
Once again, since $ e^{i\phi}$ always appears in the form ``$e^{i\phi} a$", 
it would strictly speaking be incorrect to interpret these as coordinates on a 
supermanifold.

\paragraph{Non-chiral $\mathcal{H}_\L \times \mathcal{H}_\R\,$.} 
In coordinates, the flag conditions on 
\begin{align}
    (Z^{\hat\alpha a }, \tilde Z^a{}_{\hat\alpha}) = 
    \left(
    \begin{bmatrix}
        \omega^A & \hat\pi^A \\
        \pi_{A'} & 0_{A'} \\
        \theta^I & \eta^I
    \end{bmatrix}\;,
    \begin{bmatrix}
         \tilde\pi_A & \tilde\omega^{A'} & -\tilde\theta_I \\
        0_A & \tilde{\hat\pi}^{A'} & -\tilde\eta_I
    \end{bmatrix}  \right) \in \mathcal{H}_\L \times \mathcal{H}_\R
\end{align} 
are realised by 
\begin{subequations}
\begin{align}\label{flag H nonchiral1}
    \tilde{\eta}^I\,\eta_I &= 0\;,\\
    \tilde{\hat\pi}^{A'}\,\pi_{A'} - \tilde{\eta}_I\theta^I &= 0\;,
    \label{flag H nonchiral2}\\
    \tilde{\pi}_A\,\omega^A + \tilde{\omega}^{A'}\,\pi_{A'}-\tilde\theta^I\,\theta_I &=0\;,\label{flag H nonchiral3}\\
    \tilde{\pi}_A\,\hat{\pi}^A - \tilde{\theta}^I\,\eta_I &= 0\;.
    \label{flag H nonchiral4}
\end{align}
\end{subequations}
The first condition can be solved by 
$\tilde{\eta}=\gamma\,\eta\,$,
 where $\gamma\in\mathbb{C}\,$.
As in the real case, we use a decomposition $\eta = \eta_A \hat\pi^A$.
Similarly, we introduce a decomposition $\tilde\eta =  \tilde{\hat\pi}^{A'}\tilde\eta_{A'}$ 
and to keep satisfied the first reality condition we ask $\eta_A \tilde\eta_{A'} = 0\,$;
this is always possible by a similar argument as employed in the chiral case.
The equation \eqref{flag H nonchiral4} can therefore be written as 
\begin{equation}
	(\tilde\pi_A-\tilde\theta \eta_A )\hat\pi^A = 0
\end{equation}
which can be solved, because $\hat\pi^A$ has a non vanishing body, as
\begin{equation}
	\hat\pi^A = \alpha(\tilde\pi^A - \tilde\theta\eta^A)\;,
 \quad \alpha\in\mathbb{C}\;.
\end{equation}
Similarly,  the equation \eqref{flag H nonchiral2} can be written 
\begin{equation}
	\tilde{\hat\pi}^{A'}(\pi_{A'} - \tilde\eta_{A'} \theta) = 0\;,
\end{equation}
which gives
\begin{equation}
	\tilde{\hat\pi}^{A'} = \tilde{\alpha}(\pi^{A'} - \tilde\eta^{A'} \theta)\;,
 \quad \tilde{\alpha}\in \mathbb{C}\;.
\end{equation}
We have then 
\begin{align}
    (Z^{\hat\alpha a }, \tilde Z^a{}_{\hat\alpha}) = 
    \left(
    \begin{bmatrix}
        \omega^A &~ \tilde\pi^A - \tilde\theta\eta^A\\
        \pi_{A'} &~ 0_{A'} \\
        \theta &~ \eta_A(\tilde\pi^A - \tilde\theta\eta^A)
    \end{bmatrix}\;,
    \begin{bmatrix}
         \tilde\pi_A & ~~\tilde\omega^{A'} &~~ -\tilde\theta \\
        0_A &~~ \pi^{A'}-\tilde\eta^{A'} \theta &~~ -
        (\pi^{A'}-\tilde{\eta}^{A'}\theta)\tilde{\eta}_{A'}
    \end{bmatrix}  \right)\;.
\end{align} 
Introducing $\widetilde\Theta = \tilde\theta + \tilde\eta^{A'}\tilde\omega_{A'}$
and $\Theta = \theta + \eta^{A}\omega_{A}$, we can write
\begin{equation}
	 (Z^{\hat\alpha a }, \tilde Z^a{}_{\hat\alpha}) = 
	  \left(
	 \begin{bmatrix}
		\omega^A &~ \tilde{\pi}^A
	- \widetilde{\Theta}\eta^{A} \\
		\pi_{A'} &~ 0 \\
		\Theta - \eta^{A}\omega_{A} &~ 
     \eta_A (\tilde{\pi}^A
	- \widetilde{\Theta}\eta^{A})
	\end{bmatrix}\;,
    \begin{bmatrix}
         \tilde\pi_A &~~ \tilde\omega^{A'} &~~ -\widetilde\Theta + \tilde\eta^{A'}\tilde\omega_{A'} \\
        0_A &~~ (\pi^{A'} - \tilde\eta^{A'} \Theta) &~~ 
        -(\pi^{A'}-\tilde{\eta}^{A'}\Theta)\tilde{\eta}_{A'}
    \end{bmatrix}  \right).
\end{equation}
The last condition to be solved, equation \eqref{flag H nonchiral3}, can now be expressed as 
\begin{align}\label{non chiral H : reality condition on u}
	\tilde{\pi}_A\,\omega^A + \tilde{\omega}^{A'}\,\pi_{A'}-\widetilde\Theta\,\Theta + \widetilde\Theta\eta^A\omega_A + \tilde\eta^{A'}\tilde\omega_{A'}\Theta &=0\;  \nonumber\\ \iff 
	(\tilde{\pi}_A\, - \widetilde\Theta\eta_A)\omega^A  + \tilde\omega^{A'}(\pi_{A'} - \tilde\eta_{A'}\Theta) &= \widetilde\Theta\,\Theta \;.
\end{align}
If we define the quantities $(u_+,u_-)$ via 
$iu_+ := (\tilde{\pi}_A\, - \widetilde\Theta\eta_A)\omega^A$ and 
$-iu_- :=  \tilde\omega^{A'}(\pi_{A'} - \tilde\eta_{A'}\Theta)$, 
then \eqref{non chiral H : reality condition on u} is equivalent to
\begin{equation}\label{eqn: H flag}
	iu_+ - iu_- =  \widetilde\Theta\,\Theta \;.
\end{equation}
Setting 
\begin{align}
    u_+ &= u_l - \frac{i}{2}\,\widetilde\Theta \Theta\;, &
    u_- &= u_r + \frac{i}{2}\,\widetilde\Theta \Theta\;,
\end{align}
one solves the condition \eqref{eqn: H flag} with $u_l = u_r=u\,$.
A parametrisation of $\mathcal{H}_\L \times \mathcal{H}_\R$ is then given by 
$(u, [\pi_{A'}], [\tilde{\pi}_A], \gamma, \Theta, \widetilde\Theta, \eta )$ 
where $\gamma\in\mathbb{C}\,$.
\paragraph{Reality conditions.}

 The real points are obtained
by imposing on the points $$\{(Z^{\hat\alpha b}, \widetilde{Z}_{\hat\alpha}{}^b)
~|~ {\widetilde{Z}}_{\hat{\alpha}}{}^b Z^{\hat{\alpha}a}= 0\}$$ 
of complex non-chiral $\mathcal{H}_\L \times \mathcal{H}_\R$ the reality conditions 
$\widetilde{Z}_{\hat\alpha}{}^b = 
h_{\overline{\hat\alpha}\hat\alpha}\overline{Z}^{\overline{\hat\alpha} b} $, 
where the complex structure $h_{\overline{\hat\alpha} \beta}$ 
is given in \eqref{eqn: metric}.
This gives
$\tilde{\omega}^{A'} = \overline{\omega}^{A'}$, $\tilde{\pi}_A = \overline{\pi}_A\,$, $\tilde{\hat\pi}^{A'} = {\overline{\hat\pi}}^{A'}\,$, $\tilde{\theta}^{I} = \overline{\theta}{}^{I}\,$, $\tilde{\eta}^{I} = \overline{\eta}{}^{I}$, which implies that 
$u\in\mathbb{R}\,$, 
$\gamma=e^{-2i\phi}$, and $\widetilde\Theta = \overline{\Theta} \,$.

\subsubsection{Fermionic subspace \texorpdfstring{$\I$}{}}

The last invariant subspace that we will consider are points of the form
\begin{equation}\label{Fermionic space-like infinity}
	Z^{\hat\alpha b} = 
	\begin{bmatrix}
		\omega ^{A b} \\
		{0_{A'}}^{b} \\
		\theta^{Ib}\neq 0
	\end{bmatrix}\;,
\end{equation}
i.e. such that $\pi_{A'}{}^b = 0$ but $\theta^{Ib}\neq 0$ . These points, which we collectively refer to as $\I$, therefore appear to be a sort of 
fermionic extension of time/space-like infinity $\iota :=\left\{ I^{\hat{\alpha}b}\right\}$ -- with this particular isolated point here given by $\theta^{Ib} = 0^{Ib}$.

In order for \eqref{Fermionic space-like infinity} to define a plan $\omega^{Ab}$ must be invertible, we can therefore always use the 
$\text{GL}(2,\IC)$ freedom to set $\omega^{Ab} = \delta^{Ab}$ and we are left with the complex coordinates $\theta^{Ib}\in \IC ^{0\vert 2\mathcal{N}}$. 

The super Poincar\'e group does not act transitively on this set 
\begin{equation}
	\begin{pmatrix}
        {M^B}_A & iT^{BA'} & {Q^B}_I \\
        0 & -{\widetilde{M}}_{B'}{}^{A'} & 0 \\
        0 & {\widetilde{Q}}^{JA'} & R^J{}_I
    \end{pmatrix}  	\cdot
	\begin{bmatrix}
		\delta ^{A b} \\
		{0_{A'}}^{b} \\
		\theta^{Ib}
	\end{bmatrix} = 
	\begin{bmatrix}
		{M^B}_A \delta^{Ab} + Q^B{}_I\theta^{Ib} \\
		0_{B'}{}^b \\
		R^J{}_I \theta^{Ib}
	\end{bmatrix} \;.
\end{equation}
However, it clearly stabilises it. There are two different case depending on whether $\theta^{I1}$ and $\theta^{I2}$ are linearly independent or not. Similarly to the case of $\mathcal{H}$, it is not an homogeneous space for the super Poincar\'e group and the orbits are not supermanifolds.

\paragraph{Reality condition.}
Denoting $(\theta^I{}^1,\theta^I{}^2) = (\theta^{I}, \eta^{I})$, we are going to impose the following reality conditions: 
\begin{align*}
	{(Z^1)}^2 = 0 & \Rightarrow \overline{\theta}^{{I}}\delta_{IJ}\theta^{J} = 0\;, \\
	{(Z^2)}^2 = 0 & \Rightarrow \overline{\eta}^{{I}}\delta_{IJ}\eta^{J} = 0\;, \\
	{Z^1 \cdot Z^2} = 0 & \Rightarrow \overline{\theta}_{J}\eta^{J} = 0\;.
\end{align*}
As explained for $\mathcal{H}$, these conditions do not define a submanifold. 
If $\mathcal{N}=1$, the first reality conditions impose that $\theta$ can be 
rewritten as $\theta = e^{i \phi}a$, with $a\in \IR_a$ and $ \phi \in \IR_a$.
The remaining reality conditions then give that $\eta = e^{i \psi} a$ with  $\psi \in \IR_c$. 
A parametrisation of $\I$ is then given by $(a, e^{i \phi}, e^{i \psi}) \in \IR_a \times S^1 \times S^1$.

\subsubsection{Non-chiral invariant subspaces at the boundary}

As explained in section \ref{subsec: complex case}, elements of $\CM= F(2\vert 0, 2\vert \mathcal{N}, \IC^{4\vert \mathcal{N}})$ 
are obtained as pairs $(Z^{\hat\alpha b}, {\widetilde Z}_{\hat \alpha}{}^c)$ 
of bi-supertwistors and dual bi-supertwistors satisfying 
\begin{equation}\label{eqn: flag}
	{\widetilde Z}_{\hat\alpha}{}^b Z^{\hat\alpha c} = 0\;.
\end{equation}
Non-chiral orbits will therefore be pairs of chiral orbits satisfying the condition \eqref{eqn: flag}.

One can show that the only admissible orbits under this condition are : 
$\mathscr{I}_\L \times \mathscr{I}_\R$, $\mathcal{H}_\L \times \mathcal{H}_\R$, 
$\iota^{0\vert 2\cal{N}}_\L \times \iota^{0\vert 2\cal{N}}_\R$, $\{\iota_\L \} \times \{\iota_\R \}$ and $\{\iota_\L \} \times \iota^{0\vert 2\cal{N}}_\R$.
In the cases where it applies, the stabilisers can be derived by doing the 
intersection of the chiral ones.

\section{Super cuts}

In this section we will investigate how super null cones intersect with the different 
boundaries  that we previously introduced. In particular we will see that 
$\mathscr{I}$ and $\mathcal{H}$ are generated by null supergeodesics respectively 
emanating from $\iota$ and $\I$. In order to achieve this we will need the 
realisation of null supergeodesics in terms of superambitwistors.

\subsection{Superambitwistor space}

Here we review elements of superambitwistor geometry, 
see e.g. \cite{manin_gauge_1988}.

The super ambitwistor space $\mathbb{A}$ is defined as the flag manifold 
$\mathbb{A}  := F(1|0, 3|\mathcal{N} , \mathbb{C}^{4|\mathcal{N}})$. 
Consequently we have the double fibration picture :
\begin{center}
\begin{tikzpicture}
\node {$F(1|0, 2|0, 2|\mathcal{N}, 3|\mathcal{N}, \mathbb{C}^{4\vert \mathcal{N}})$}[sibling distance = 5cm]
    child {node {$\mathbb{A} =  F(1|0, 3|\mathcal{N}, \mathbb{C}^{4|\mathcal{N}})$} edge from parent node [left] {$\pi_1$}}
    child {node {$\overline{M} = F(2|0, 2|\mathcal{N}, \mathbb{C}^{4|\mathcal{N}})$~~.} edge from parent node [right] {$\pi_2$}};
\end{tikzpicture}	
\end{center}

In practice elements of $\mathbb{A} = F(1|0, 3|\mathcal{N} , \mathbb{C}^{4\vert \mathcal{N}})$ are given by  pairs $(Z^{\hat\alpha}, \widetilde{Z}_{\hat\alpha})$ 
of twistors (and dual twistors) satifying $\widetilde{Z}_{\hat\alpha} Z^{\hat\alpha}=0$. Similarly, elements of $F(1|0, 2|0, 2|\mathcal{N}, 3|\mathcal{N} ,\mathbb{C}^{4\vert \mathcal{N}})$ are given by quadriplets $(Z^{\hat\alpha}{}^{a} \pi_{a}, Z^{\hat\alpha}{}^a, \widetilde{Z}_{\hat\alpha}{}^{a}, \widetilde{Z}_{\hat\alpha}{}^{a} \tilde{\pi}_{a}) $.
\begin{center}
\begin{tikzpicture}
\node {($ Z^{\hat\alpha}{}^{a} \pi_{a}, Z^{\hat\alpha}{}^a, \widetilde{Z}_{\hat\alpha}{}^a, \widetilde{Z}_{\hat\alpha}{}^a  \tilde{\pi}_{a}$)}[sibling distance = 3.5cm]
    child {node {($Z^{\hat\alpha}{}^{a} \pi_{a} , \widetilde{Z}_{\hat\alpha}{}^a \tilde{\pi}_{a}$)} edge from parent node [left] {$\pi_1$}}
    child {node {($Z^{\hat\alpha}{}^a, \widetilde{Z}_{\hat\alpha}{}^a$)~~,} edge from parent node [right] {$\pi_2$}};
\end{tikzpicture}	
\end{center}
from which one clearly sees e.g. that a point
 $(Z^{\hat\alpha}{}^a , \widetilde{Z}_{\hat\alpha}{}^a)$ in Minkowski superspace 
 gives an embedding of $\mathbb{CP}^1 \times \mathbb{CP}^1$ into $\mathbb{A}$ : 
 $([\pi_{a}], [\tilde{\pi}_{a}] ) \mapsto (Z^{\hat\alpha}{}^{a} \pi_{a} , \widetilde{Z}_{\hat\alpha}{}^a  \tilde{\pi}_{a})$. 

\subsubsection{Null supergeodesics}

In the reverse sense, a point of the superambitwistor space
\begin{align}\label{SuperAmbitwistor: SuperAmbitwistor space}
    (Z^{\hat\alpha}, \widetilde{Z}_{\hat\alpha}) &= (\begin{bmatrix} \omega^{A} \\ \pi_{A'} \\ \theta^I
    \end{bmatrix}, 
    \bm{[} \;\tilde{\pi}_{A} \quad \tilde{\omega}^{A'} \quad -\tilde{\theta}_I
    \;\bm{]} ) \quad\mbox{s.t.}\quad \widetilde{Z}_{\hat\alpha} Z^{\hat\alpha} =\tilde{\pi}_{A} \omega^{A} + \tilde{\omega}^{A'}\pi_{A'} - \tilde{\theta}_I\theta^I =0\;,
\end{align}
defines a null supergeodesic in the (conformal compactification of) 
complexified Minkowski space. To see what these are, let us look what the 
condition is for one point of Minkowski to be on this null supergeodesic. 
For that, one works in the local chart defining Minkowski superspace:
\begin{equation}
    (Z^{\hat\alpha}{}^b, \widetilde{Z}_{\hat\alpha}{}^b )
    = (\begin{bmatrix} iX_+^{AB'} \\ 
    \delta_{A'}{}^{B'}\\ \theta^I{}^{B'}\end{bmatrix} ,  
    \bm{[}\; \delta_{A}{}^B \quad -iX_-^{A'B} \quad 
    -\tilde{\theta}_I{}^{B}\;\bm{]})
\end{equation} 
with
\begin{align*}
    X^{AB'}_+ &= x^{AB'} - \frac{i}{2}\,\tilde{\theta}_I{}^A\theta^{IB'}, &
    X^{A'B}_- &= x^{A'B} + \frac{i}{2}\,\tilde{\theta}_I{}^B\theta^{IA'}\,.
\end{align*}
This spacetime point is part of the null geodesics defined by
the ambitwistor \eqref{SuperAmbitwistor: SuperAmbitwistor space} if and only if it is in the image of this point by the double fibration:
\begin{align}\label{eqn: ambitwistor cond}
    Z^{\hat\alpha} &= Z^{\hat\alpha}{}^{b}\pi_b\;, &
    \widetilde{Z}_{\hat\alpha} & =  \widetilde{Z}_{\hat\alpha}{}^{b}\tilde{\pi}_b\;,
\end{align}
 i.e. if the point satisfies
\begin{align}
        \omega^A &  = iX^{AB'}_+ \pi_{B'}\;, & \theta^I &= \theta^{IB'}\pi_{B}'\;, \\
        \tilde{\omega}^{B'} & = -iX_-^{AB'} \tilde{\pi}_{A}\;,  &\tilde{\theta}_I &= \tilde{\theta}_I{}^{B}\tilde{\pi}_{B}\;.
\end{align}
Solutions to these equations can be parametrised by $(\epsilon,  \epsilon^I , \tilde{\epsilon}^I) \in \mathbb{C}_c \times \mathbb{C}_a^{\mathcal{N}}\times\mathbb{C}_a^{\mathcal{N}} $ as follows: 
\begin{align}
iX^{BB'}_+ &= iX_0^{BB'} + \tilde{\theta}_0{}_I{}^B \epsilon^I \pi^{B'} +(i \epsilon -\frac{1}{2} \epsilon^I \tilde{\epsilon}_I)\pi^{B'}\tilde{\pi}^{B}, &  \theta^{IB'} &= \theta_0^{IB'} + \epsilon^I\pi^{B'}, \\
iX^{B'B}_- &= i\tilde{X}_0^{BB'} +\theta^{IB'}_0\tilde{\epsilon}_I \tilde{\pi}^B   +(i\epsilon +\frac{1}{2} \epsilon^I \tilde{\epsilon}_I)\pi^{B'}\tilde{\pi}^{B},&
 \tilde{\theta}_I{}^{B} &= \tilde{\theta}_0{}_I{}^{B} + \tilde{\epsilon}_I\tilde{\pi}^{B},
\end{align}
where $(X_0^{AA'}, \tilde{X}_0^{A'A}, \theta_0^{IA'}, \tilde{\theta}_{0I}{}^A)$ is any solution of the equations. Equivalently
\begin{align}
x^{AA'} &= x_0^{AA'} -\frac{i}{2}\left( \pi^{A'} \tilde{\theta}_0{}_I{}^A \epsilon^I + \theta^{IA'}_0 \tilde{\pi}^A \tilde{\epsilon}_I \right)  +\epsilon \pi^{A'}\tilde{\pi}^{A},\\
\theta^{IB'} &= \theta_0^{IB'} + \epsilon^I\pi^{B'}, \\
 \tilde{\theta}_I{}^{B} &= \tilde{\theta}_0{}_I{}^{B} + \tilde{\epsilon}_I\tilde{\pi}^{B}.
\end{align}
where $X^{AB'}_0 = x^{AB'}_0 - \frac{i}{2}\,\tilde{\theta}_0{}_I{}^A\theta_0^{IB'}\,$, 
$\tilde{X}^{B'A}_0 = x^{AB'}_0 +\frac{i}{2}\,\tilde{\theta}_0{}_I{}^A\theta_0^{IB'}\,$. 

Keeping all the other parameters fixed in the above equations, 
$(\epsilon,  \epsilon^I , \tilde{\epsilon}^I) \in \mathbb{C}_c \times \mathbb{C}_a^{\mathcal{N}}\times\mathbb{C}_a^{\mathcal{N}}$ parametrise 
the super null geodesic emanating from $(Z_0{}^{\hat\alpha a}, \widetilde{Z}_0{}_{\hat\alpha}{}^a)$ and in the direction $(\pi^a, \tilde{\pi}^a)$. 
Varying $(\pi^a, \tilde{\pi}^a)$ allows to span the whole supernull cone emanating 
from $(Z_0{}^{\hat\alpha a}, \widetilde{Z}_0{}_{\hat\alpha}{}^a)$.

\subsection{Cuts along super \texorpdfstring{$\mathscr{I}$}{}}

\subsubsection{\texorpdfstring{$\mathscr{I}$}{} as the union of super null lines emanating from infinity}

We are here interested in the super null cone emanating from \eqref{eqn: super null infinity twistor}
\begin{align}\label{SuperGoodCuts: super null cone at infinity2}
   \{\iota \} =  \left( I^{\hat\alpha}{}^b = \begin{bmatrix}
        \delta^{Ab} \\ 0 \\   0
    \end{bmatrix},   \tilde{I}_{\hat\alpha}{}^b = 
        \bm{[} \;0 \quad \delta^{A'}{}^b \quad 0\; \bm{]}
    \right).
\end{align}
Each of the genenerators of this super null cone corresponds to a superambitwistor
  \begin{equation}\label{SuperGoodCuts: super null cone at infinity}
     (I^{\hat\alpha}{}^b \tilde{\lambda}_{b}, \tilde{I}_{\hat\alpha}{}^{b} {\lambda}_{b})\;.
  \end{equation}

At infinity, these null cones intersect several different orbits of the super Poincaré group.

The intersection of the super null cone emanating from $\{\iota\}$ 
with super $\mathscr{I}$ is composed of points $(Z^{\alpha}{}^b,\widetilde{Z}_{\alpha}{}^b )$ 
of the form:
\begin{align}
    Z^{\hat\alpha}{}^b &= \begin{bmatrix}
        \omega^A & \tilde{\lambda}^A \\ 
        {\lambda}_{A'} & 0 \\   \theta^I & 0
    \end{bmatrix},  &  \widetilde{Z}_{\hat\alpha}{}^b &= \begin{bmatrix}
        \tilde{\lambda}_{A} & \tilde{\omega}^{A'} & -\tilde{\theta}_I  \\ 0 & \lambda^{A'}  & 0
    \end{bmatrix}
\end{align}
with $\tilde\lambda_A\omega^A  + \tilde{\omega}^{A'}\lambda_{A'} - \tilde{\theta}^I \theta_I =0\,$. 

Introducing $iu_+  = \tilde\lambda_A\omega^A$ and $-iu_-  = \tilde{\omega}^{A'}\lambda_{A'}$, this last equation can be solved as
\begin{align}
    u_{+} &= u + \frac{i}{2}\, \tilde{\theta}^I\theta_I\;, & u_{-} &= u - \frac{i}{2} 
    \,\tilde{\theta}^I\theta_I\;,
\end{align} with $u \in \IR$.
These points all lie in $\mathscr{I}$ and any point in  $\mathscr{I}$ can be obtained in this way: in the coordinate system $(u ,  [\lambda_{A'}] , [\tilde{\lambda}_{A}],  \theta^I, \tilde{\theta}^I)$ the corresponding super null line is obtained by varying $u$, $\theta^I$, $\tilde{\theta}^I$ while keeping $[\lambda_{A'}]$ and $[\tilde{\lambda}_{A}]$ fixed.

\subsubsection{Cuts at \texorpdfstring{$\mathscr{I}$}{} emanating from a point at finite distance}

We consider the super null cone emanating from a point at finite distance
\begin{align}\label{SuperGoodCuts: minkoswski coordinates 2}
	W^{\hat\alpha b} &= \begin{bmatrix}
		iX^{AB'}_+ \\
		\delta_{A'}{}^{B'} \\
		\theta ^{IB'}
	\end{bmatrix}, & \widetilde{W}_{\hat\alpha b} &= \begin{bmatrix}
		\delta_{A}{}^{B} &
		 -i X^{A'B}_- &
		-\tilde{\theta}^{IB}
	\end{bmatrix}
\end{align}
with $X^{B'C}_{\pm} = x^{B'C} \mp \frac{i}{2}\,\tilde{\theta}_I{}^C\theta^{IB'}$.
Varying $(\lambda_{b}, \tilde{\lambda}_{b})$, the super ambitwistors
  \begin{equation}\label{SuperGoodCuts: super null cone}
     (W^{\hat\alpha}{}^b \lambda_{b}, \widetilde{W}_{\hat\alpha}{}^b \tilde\lambda_{b})
  \end{equation}
 then correspond to the super null cone passing through this point. 
 The intersection of this super null cone with $\mathscr{I}$ is the  cut 
 \begin{align}
    &&(\lambda^a , \tilde{\lambda}^a) &&\mapsto && \left( \begin{bmatrix} i X^{AB'}_+\lambda_{B'} & \tilde{\lambda}^A \\ \lambda_{A'} & 0 \\ \theta^{IB'}\lambda_{B'}  & 0
 \end{bmatrix} , 
 \begin{bmatrix}  \tilde{\lambda}_{A} & -i {X}^{A'B}_- \tilde{\lambda}_{B} &  -\tilde{\theta}^{IB}\tilde{\lambda}_B \\  0 & \lambda^{A'}  & 0
 \end{bmatrix} \right).
\end{align}
\paragraph{Non-chiral representation of the cuts.}
In the (complex) \emph{non-chiral} coordinate system \\
$(u ,  [\pi_{A'}] , [\tilde{\pi}_{A}],  \theta^I, \tilde{\theta}^I)$  for $\mathscr{I}$, 
the cuts of $\mathscr{I}$ by super null cone are:
 \begin{align}
    &&(\lambda^a , \tilde{\lambda}^a) &&\mapsto && 
    \left( x^{AA'} \lambda_{A'}\tilde{\lambda}_{A},   [\lambda_{A'}] , [\tilde{\lambda}_{A}],  \theta^{IB'}\lambda_{B'} , \tilde{\theta}^{IB}\tilde{\lambda}_B \right).
\end{align}

\paragraph{Chiral representation of the cuts.}
In the (complex) \emph{chiral}  left coordinate system 
\\ $(u_+ ,  [\pi_{A'}] , [\hat{\pi}_{A}],  \theta^I)$  for $\mathscr{I}$, the cuts of $\mathscr{I}$ by super null cone are:
 \begin{align}
    &&(\lambda^a , \tilde{\lambda}^a) &&\mapsto && 
    \left( X^{AA'}_+ \lambda_{A'}\tilde{\lambda}_{A},   [\lambda_{A'}] , [\tilde{\lambda}_{A}],  \theta^{IB'}\lambda_{B'}\right).
\end{align}

\subsection{Cuts along \texorpdfstring{$\mathcal{H}$}{}} 

\subsubsection{\texorpdfstring{$\mathcal{H}$}{} as the union of super null lines emanating from \texorpdfstring{$\I$}{}}

 Here we discuss the intersection with $\mathcal{H}$ of a super null cone 
 emanating from a point of $\I\,$ which has the form
 \begin{align}
	W^{\hat\alpha b} &= \begin{bmatrix}
		\delta^{Ab} \\
		0\\
		\eta ^{Ib}
	\end{bmatrix}, & \widetilde{W}_{\hat\alpha}{}^b &= \begin{bmatrix}
		0 &
		 \delta^{A'b} &
		-\tilde{\eta}^{Ib}
	\end{bmatrix},
\end{align}
with $\tilde{\eta}^{IB'}\eta_I{}^B =0$.
This intersection is given by points of the form 
\begin{align}\label{SuperGoodCuts: generator at H}
Z^{\hat\alpha b} &= \begin{bmatrix}
        \omega^A & \tilde\lambda^{A} \\ \tilde{\eta}_I{}_{A'}\theta^I  +  \tilde\alpha{\lambda}_{A'} & 0 \\   \theta^I & \eta^{IB} \tilde\lambda_{B}
    \end{bmatrix},  &  \tilde{Z}_{\hat\alpha}{}_b &= \begin{bmatrix}
        \tilde{\theta}^I \eta_I{}_{A} + \alpha\tilde\lambda_{A} &\; \tilde{\omega}^{A'} &  -\tilde{\theta}^I \\
        0 & {\lambda}^{A'}  & -\tilde{\eta}_{IB'}{\lambda}^{B'}
    \end{bmatrix}.
\end{align}
with the constraint
$\alpha \tilde\lambda_A\omega^A 
+\tilde\alpha \lambda_{A'}\tilde\omega^{A'}
= (\tilde\theta^I-\tilde\eta^I_{A}\tilde\omega^{A'})(\theta_I - \eta_{IA}\omega^A)$.

Introducing $iu_+  = \alpha\tilde \lambda_A\omega^A$, 
$-iu_-  = \tilde\alpha\lambda_{A'}\tilde{\omega}^{A'}$, 
and $\Theta^I = \theta^I - \eta^I{}_A\omega^A$ , $\widetilde\Theta^I = \tilde\theta^I - \tilde\eta^I{}_{A'}\tilde\omega^{A'}$, the above constraint equation 
can be solved by $iu_+ - iu_- = \widetilde\Theta_I\Theta^I$, i.e., 
\begin{align}
    u_{+} &= u - \frac{i}{2} \widetilde{\Theta}_I\Theta^I, & u_{-} &= u + \frac{i}{2} \widetilde{\Theta}_I\Theta^I\;,\qquad u\in \mathbb{C}\;.
\end{align}
First, these points \eqref{SuperGoodCuts: generator at H} 
all lie in $\mathcal{H}$: in the coordinate system 
  $(u , [\lambda_{A'}], [\tilde\lambda_{A}] ,  \Theta^I, \widetilde{\Theta}^I, \eta^I, \tilde{\eta}^I )$
on $\mathcal{H}$ the corresponding super null line is obtained by varying $u$, $\Theta^I$, $\widetilde{\Theta}^I$ while keeping $\eta^I =  \eta^{IB} \tilde\lambda_{B}$, $\tilde{\eta}^I =  \tilde{\eta}^{IB'} \lambda_{B'}$  , $[\tilde\lambda_A]$ and $[\lambda_{A'}]$ fixed.

Second, any point in $\mathcal{H}$ can be obtained in this way, since we showed 
in \ref{subsubsec: H} that $\eta^I$ and $\tilde\eta_I$ can always be written in a 
factorised form $\eta^{IB}\,\tilde\lambda_B$
and $\tilde{\eta}_{IB'}\lambda^{B'}\,$, respectively. 

\subsubsection{Cuts at \texorpdfstring{$\mathcal{H}$}{} emanating from a point at finite distance}

In this section we consider the super null cone emanating from a point at finite distance
\eqref{SuperGoodCuts: minkoswski coordinates 2}.

The intersection of this super null cone with $\mathcal{H}$ is given by points of the form 
\begin{align}
Z^{\hat\alpha a} &=\begin{bmatrix} i X^{AA'}_+\lambda_{A'} & 
~\tilde\lambda^B (\delta^A_B-\eta^A_I\tilde{\theta}^I_B) \\ \lambda_{A'} & 0 \\ \theta^{IA'}\lambda_{A'}  
& \tilde{\lambda}^B(\eta_{IB}-\eta_{IC}\eta^{C}_J\tilde{\theta}^J_B)
 \end{bmatrix}\;,
\end{align}
\begin{align}
  \widetilde{Z}^b{}_{\hat\alpha} &=\begin{bmatrix} 
  \tilde\lambda_A & -iX_-^{A'B}\tilde\lambda_B & -\tilde\theta_I{}^B\tilde\lambda_B \\
  0 &~ \lambda^{B'}(\delta^{A'}_{B'}+\tilde\eta^{A'}_I \theta^I_{B'})  
  & ~-\lambda^{B'}(\tilde\eta^I_{B'}+\tilde\eta^I_{C'}\tilde\eta^{C'}_J{\theta}^J_{B'})
 \end{bmatrix}\;,
\end{align}
where $\tilde\eta^I{}_A \eta_{IA'} = 0$ and 
$X^{B'C}_{\pm} = x^{B'C} \mp \frac{i}{2}\,\tilde{\theta}_I{}^C\theta^{IB'}$.
In terms of the variables $\Theta^I, {\widetilde\Theta}_I$ defined in section
\ref{subsubsec: H}, here taking the values 
$\Theta^I = \theta^{IA'}\lambda_{A'}-i\eta^I_AX_+^{AB'}\lambda_{B'}\,$, 
$\widetilde\Theta_I = \tilde\theta^{A}_I\tilde\lambda_{A}+i\tilde\eta_{A'I}X_-^{A'B}
\tilde\lambda_{B}\,$, 
we have 
\begin{align}
Z^{\hat\alpha a} &
=\begin{bmatrix} i X^{AA'}_+\lambda_{A'} 
& 
~\tilde\lambda^A - \widetilde\Theta^I\eta^A_I \\ \lambda_{A'} & 0 
\\ \Theta^{I}-i\eta^I_AX_+^{AB'}\lambda_{B'}  & 
\eta_{IA}(\tilde\lambda^A-{\widetilde\Theta}^J\eta^A_J)
 \end{bmatrix}, &
  \widetilde{Z}^b{}_{\hat\alpha} &=\begin{bmatrix} 
  \tilde\lambda_A & -iX_-^{A'B}\tilde\lambda_B 
  & -\widetilde\Theta_I+i X_-^{A'B}\tilde{\eta}_{A'I}\tilde\lambda_B \\
  0 &~ \lambda^{A'}-\tilde\eta^{A'}_I \Theta^I  
  & ~-(\lambda^{B'}-\tilde\eta_J^{B'}{\Theta}^J){\tilde\eta}^I_{B'}
 \end{bmatrix}.
\end{align}

\paragraph{Non-chiral representation of the cuts.}
In the complex non-chiral coordinate system \\
$(u, [\pi_{A'}], [\tilde{\pi}_A], \Theta^I, \widetilde\Theta^I, \eta^I , \tilde \eta^I )$ 
on $\mathcal{H}_\L \times \mathcal{H}_\R\,$, such that $\tilde\eta_I \,\eta^I=0\,$, 
the cuts of $\mathcal{H}_\L \times \mathcal{H}_\R$ by super null cone are:
\begin{align}
    &(\lambda^{A'} , \tilde{\lambda}^{A}, \eta^I_{A}, \tilde\eta^I_{A'})  
    \mapsto  \\
    &\left( u\,, \ [\lambda_{A'}] , \ [\tilde{\lambda}_{A}], 
\;[\theta^{IB'}-i\eta_A^I(x^{AB'}
-\tfrac{i}{2}\tilde\theta^A_J\theta^{JB'})]\lambda_{B'} \;,
\;\tilde{\lambda}_B[\tilde\theta^B_I + i(x^{BA'}
+\tfrac{i}{2}\tilde\theta^B_J\theta^{JA'})\tilde\eta_{A'I}]\;,
\; \eta^I \;,\; \tilde\eta_I \right)\;,\nonumber
\end{align}
where $\eta^I=\eta^I_A(\tilde\lambda^A-{\widetilde\Theta}^J\eta^A_J)\,$,
$\tilde\eta_I=(\lambda^{B'}-\tilde\eta_J^{B'}{\Theta}^J){\tilde\eta}_{IB'}\,$,
and 
\begin{align}
 u = x^{AB'}\tilde\lambda_A\lambda_{B'}
 +\tfrac{1}{2}\,\eta_{IA}x^{AB'}\lambda_{B'}\tilde{\theta}^I
 +\tfrac{1}{2}\,{\theta}^I\tilde\lambda_{B}x^{BA'}\tilde\eta_{A'I}
 +\tfrac{i}{4}\,(\eta_{IA}\tilde{\theta}^A_J)(\tilde{\theta}^I\theta^J)
 -\tfrac{i}{4}\,(\theta_J^{A'}\tilde{\eta}_{A'}^I)(\tilde{\theta}^J\theta^I)\;.  
\end{align}

\paragraph{Chiral representation of the cuts.}
In the complex \emph{chiral} left coordinate system 
\\ $(u_+, [\pi_{A'}], [\hat{\pi}^A], \Theta^I, \eta^I)$
for $\mathcal{H}_\L$, the ``cuts'' of $\mathcal{H}_\L$ by super null cone are:
\begin{align}
    (\lambda^a , \tilde{\lambda}^a,   , \eta^I_A ) &\mapsto  
    \\
    &\left(  X_+^{AB'}\lambda_{B'}\tilde\lambda_C( \delta_A^C - \tilde\theta_A^I \eta^C_I ), \   [\lambda_{A'}] , [ ( \delta^A_C +\tilde\theta_I{}^A\eta^I_C) \tilde \lambda^{C} ], \ (\theta^{IB'} - i\eta_A^I X^{AB'}_+)\lambda_{B'} , \ \eta^I_A \tilde\lambda^A \right).\nonumber
\end{align}
\vspace{2cm}
\section*{Acknowledgements}
The authors are grateful to Lionel Mason for many enlightening discussions.
We would also like to thank Juyoung Park, Stefan Prohazka and Ergin Sezgin.
The work of N.B. was partially supported 
by the F.R.S.-FNRS PDR grant T.0022.19 ``Fundamental issues in extended 
gravitational theories". 
N.B. wants to thank the hospitality of the Institut Denis Poisson, Tours, 
where the work was finalised.  
N.P. would like to thank the Fonds de 
la Recherche Scientifique$\,$--FNRS for financial support.

\newpage
\appendix

\section{Notations and conventions}
\label{app:notations}

The notation $[x]$ is used to express that $x$ is a choice of representative 
in an equivalence class.
A matrix $N$ of $sl(2,\IC)$ will be denoted $N=(N^A{}_B)\,$.
The complex conjugated matrix $\overline{N}$ will be denoted 
$\overline{N}=(\overline{N}{}^{A'}{}_{B'})\,$. 
We recall that the Pauli matrices $\vec{\sigma}$  are given by 
\begin{align}
	\sigma_1 &= \begin{pmatrix}
		0 & 1 \\
		1 & 0
	\end{pmatrix}\;,
	&
	\sigma_2 &= \begin{pmatrix}
		0 & -i \\
		i & 0
	\end{pmatrix}\;,
	&
	\sigma_3 &= \begin{pmatrix}
		1 & 0 \\
		0 & -1
	\end{pmatrix}\;.
\end{align}
In our convention, the four matrices $\sigma_a = (\mathbb{I},\vec{\sigma})$ 
carry the indices $\sigma_a^{AA'}\,$ while the four matrices 
$\tilde{\sigma}^a=(\mathbb{I},-\vec{\sigma})$ carry the indices 
$\tilde\sigma^a_{A'A} = \epsilon_{A'B'}\epsilon_{AB}\sigma^{aBB'}$.
We use conventions whereby the invariant $sl(2,\IC)$ matrices are
\begin{align}
	(\epsilon_{AB}) &= \begin{pmatrix}
		0 & 1 \\
		-1 & 0
	\end{pmatrix}\;,\quad
 & (\epsilon^{AB}) =
 \begin{pmatrix}
		0 & 1 \\
		-1 & 0
	\end{pmatrix}\;,
\end{align}
such that 
\begin{align}
    \epsilon^{AB}\,\epsilon_{CB} = \delta^A{}_C\;.
\end{align}
This corresponds to the following convention for lowering 
the indices of a (left) spinor $\omega^A\,$:
\begin{align}
    \omega_A = \omega^B\,\epsilon_{BA}\quad \Leftrightarrow\quad 
    \omega^A = \epsilon^{AB}\,\omega_B\;.
\end{align}
The same convention applies for lowering and raising the 
indices of a right spinor $\pi^{A'}\,$ using the invariant 
matrices $(\epsilon^{A'B'})$ and $(\epsilon_{A'B'})$, numerically
equal to the matrices $(\epsilon^{AB})= (\epsilon_{AB})\,$.

Given that, we have 
\begin{align}
	{(\sigma_a)}^{AA'} &= {(\mathbb{I}, \sigma_i)}^{AA'}\;, &
	{(\tilde{\sigma}_a)}_{A'A} &=  {(\mathbb{I}, -\sigma_i)}_{A'A}\;, \\
	{(\sigma_{ab})}^A{}_B &= -\frac{1}{4}{(\sigma_a\tilde{\sigma}_b - \sigma_b\tilde{\sigma}_a)}^A{}_B \;,&
	{(\tilde{\sigma}_{ab})}_{A'}{}^{B'} &= -\frac{1}{4}{(\tilde\sigma_a{\sigma}_b - \tilde\sigma_b{\sigma}_a)}_{A'}{}^{B'}\;,
\end{align}
and then, 
\begin{equation}
	\gamma_a = \begin{pmatrix}
		0 & \sigma_a \\
		\tilde{\sigma}_a & 0
	\end{pmatrix}\;,\qquad
 \gamma_5 = \begin{pmatrix}
		\mathbb{I} & 0 \\
		0 & -\mathbb{I}
	\end{pmatrix}\;.
\end{equation}

\section{The superconformal algebra}
\label{appendixSuperConf}

In this appendix we choose a basis for the superconformal algebra in terms of 
matrices and 
compute the (anti)commutation relations between the generators of this basis. 
We stick as much as possible to the conventions of \cite{buchbinder_ideas_1998}.

\begin{itemize}
	\item Translations : we can decompose $i\,t^{AB'} = -i\,b^a{(P_a)}^{AB'}$ 
        for certain coefficents $b^a \in \mathbb{R}_c$ and bosonic generators 
        $P_a\,$. 
		We fix ${(P_a)}^{AB'} = -\tfrac{1}{4}\,{(\sigma_a)}{}^{AB'}\,$. 
        The matrix of coefficients $t^{AB'}$ is indeed hermitian with this choice.
		\item Dilations and Lorentz transformations: we can decompose 
		$ m^A{}_B = \delta\, D^A{}_B + \frac{i}{2} \lambda_{ab}{(J^{ab})}^A{}_B$, 
        with $\delta\,,$ $\lambda_{ab} \in \IR_c\,$. 
        We fix $D^A{}_B = \frac{1}{2}\delta^A{}_B\,$
        and ${(J_{ab})}^A{}_B = -i{(\sigma_{ab})}^A{}_B\,$.
        \item Special conformal transformations : we can decompose $-ik_{A'B} = -if^a{(K_a)}_{A'B}$, where $f^a \in \mathbb{R}_c$. 
        We fix ${(K_a)}_{A'B} = \frac{1}{4}{({\tilde{\sigma}}_a)}{}_{A'B}\,$.
         \item Supercharges : we can decompose 
        $q^A{}_I = i\,\xi_B{}_J {(Q^B{}^J)}{}^A{}_I $ where 
        ${(Q^B{}^J)}{}^A{}_I = \epsilon^{AB}\,\delta^J{}_I\,$. Similarly,  
        $\overline{q}^{IA'} = i\,\overline{\xi}_{B'}^J \,
        (\overline{Q}{}^{B'}_J){}^{IA'}$
        where we have 
        $(\overline{Q}{}^{B'}_J){}^{IA'}=\epsilon^{A'B'}\,\delta^I_J\,$.
        These generators $Q$ and $\overline{Q}$ are odd.
        \item Super special conformal transformations: 
        we decompose $s^I{}_B = i\,\eta_A{}^J({S^A{}_J)}{}^I{}_B$ where 
        the matrix elements $({S^A{}_J)}{}^I{}_B = \delta^A_B\,\delta^I_J\quad$ 
        and $\eta\in\mathbb{R}_a\,$. Similarly, 
        $\overline{s}{}_{A'J} = i\, \overline{\eta}{}_{B'I}\,
        (\overline{S}{}^{B'I}){}_{A'J}$ where 
        $(\overline{S}{}^{B'I}){}_{A'J}=\delta^{B'}_{A'}\,\delta^I_J\,$.
        \item $R$-symmetries : The ${\cal N}^2-1$ generators of 
        $SU({\cal N})$ 
        can be written in terms of the ${\cal N}\times {\cal N}$ matrices 
        $E_{I,J}$ such that $(E_{I,J}){}^K{}_L = \delta^K{}_I\,\delta_{JL}\,$ 
        with matrix product $E_{I,J}E_{K,L}=\delta_{JK}E_{I,L}\,$. 
        In order to give a basis of traceless hermitian matrices in the defining 
        representation of $\mathfrak{su}({\cal N})\,$, we take the union 
        ${\cal T}_D\cup {\cal T}_S\cup{\cal T}_A\,$ where 
        \begin{align}
            {\cal T}_D &= \{ E_{I,I}-E_{I+1,I+1}\;\vert\; 
            I \in\{ 1,\ldots {\cal N}-1\}\}\;,
            \nonumber \\
            {\cal T}_A & = \{ i\,(E_{I,J} - E_{J,I})
            \;\vert\; 1\leqslant I<J\leqslant {\cal N}\}\;,
            \nonumber \\
            {\cal T}_S &= \{ E_{I,J} + E_{I,J} \;\vert\; 1\leqslant I<J 
            \leqslant {\cal N}\}\;.
            \nonumber
        \end{align}
        In the first nontrivial case where ${\cal N}=2\,$, this basis reproduces the three 
        Pauli matrices. The set ${\cal T}_D$ is made of diagonal matrices, while the 
        sets ${\cal T}_A$ and ${\cal T}_S$ are made of antisymmetric and symmetric matrices, 
        respectively. The former set provides a basis of 
        $\mathfrak{so}({\cal N})\subset \mathfrak{su}({\cal N})\,$.
\end{itemize}

We then fix $\mathcal{N} = 1$ and provide the following 
matrix realisation of the superconformal algebra:
\begin{align}\nonumber
P_a&= \left(\begin{array}{cccccc}
\ & \ & \ & \ & \temp & 0\\ 
\ & \ & -\frac{1}{8}(\mathbb{I}_4 +\gamma_5)\gamma_a & \ & \temp & 0 \\
\ & \ &  & \ & \temp & 0 \\
\ & \ & \ & \ & \temp & 0 \\ \cline{1-6}
0 & 0 & 0 & 0 & \temp & 0 
\end{array}\right) =\left(\begin{array}{cccccc}
0_{2} & \ & -\tfrac{1}{4}\,{(\sigma_a)}{}^{AB'} &\ & \temp & 0\\ 
\ & \ & \ & \ & \temp & 0 \\
0_{2} & \ & 0_{2} & \ & \temp & 0 \\
\ & \ & \ & \ & \temp & 0 \\ \cline{1-6}
0 & 0 & 0 & 0 & \temp & 0 
\end{array}\right) \;,
\end{align}
\begin{align}\nonumber
D &=\left(\begin{array}{cccccc}
\tfrac{1}{2} & 0 & 0 & 0 & \temp & 0\\ 
0 & \tfrac{1}{2} & 0 & 0 & \temp & 0 \\
0 & 0 & -\tfrac{1}{2} & 0 & \temp & 0 \\
0 & 0 & 0 & -\tfrac{1}{2} & \temp & 0 \\ \cline{1-6}
0 & 0 & 0 & 0 & \temp & 0 
\end{array}\right) = \frac{1}{2}\left(\begin{array}{cccccc}
\ & \ & \ & \ & \temp & 0\\ 
\ & \ & \ & \ & \temp & 0 \\
\ & \ & \gamma_5 & \ & \temp & 0 \\
\ & \ & \ & \ & \temp & 0 \\ \cline{1-6}
0 & 0 & 0 & 0 & \temp & 0 
\end{array}\right)\;,
\end{align}
\begin{align}\nonumber
J_{ab} &= \left(\begin{array}{cccccc}
\ & \ & \ &\ & \temp & 0\\ 
\ & \ & \ & \ & \temp & 0 \\
\ & \ & -i\Sigma_{ab} & \ & \temp & 0 \\
\ & \ & \ & \ & \temp & 0 \\ \cline{1-6}
0 & 0 & 0 & 0 & \temp & 0 
\end{array}\right) = \left(\begin{array}{cccccc}
-i{(\sigma_{ab})}{}^A{}_B & \ & 0_2 &\ & \temp & 0\\ 
\ & \ & \ & \ & \temp & 0 \\
0_{2} & \ & -i{({\tilde\sigma}_{ab})}{}_{A'}{}^{B'} & \ & \temp & 0 \\
\ & \ & \ & \ & \temp & 0 \\ \cline{1-6}
0 & 0 & 0 & 0 & \temp & 0 
\end{array}\right)\;,
\end{align}
\begin{align}\nonumber
K_a &= \left(\begin{array}{cccccc}
\ & \ & \ & \ & \temp & 0\\ 
\ & \ & -\frac{1}{8}(\mathbb{I}_4 -\gamma_5)\gamma_a & \ & \temp & 0 \\
\ & \ &  & \ & \temp & 0 \\
\ & \ & \ & \ & \temp & 0 \\ \cline{1-6}
0 & 0 & 0 & 0 & \temp & 0 
\end{array}\right) = \left(\begin{array}{cccccc}
0_{2} & \ & 0_2 &\ & \temp & 0\\ 
\ & \ & \ & \ & \temp & 0 \\
-\frac{1}{4}{({\tilde{\sigma}}_a)}_{A'B} & \ & 0_{2} & \ & \temp & 0 \\
\ & \ & \ & \ & \temp & 0 \\ \cline{1-6}
0 & 0 & 0 & 0 & \temp & 0 
\end{array}\right)\;,
\end{align}
\begin{align}\nonumber
Q^1 &= \left(\begin{array}{cccccc}
\ & \ & \ & \ & \temp & 0\\ 
\ & \ & \ & \ & \temp & -1 \\
\ & \ & 0_4 & \ & \temp & 0 \\
\ & \ & \ & \ & \temp & 0 \\ \cline{1-6}
0 & 0 & 0 & 0 & \temp & 0 
\end{array}\right)\;,\ \ \ \  
Q^2 = \left(\begin{array}{cccccc}
\ & \ & \ & \ & \temp & 1\\ 
\ & \ & \ & \ & \temp & 0 \\
\ & \ & 0_4 & \ & \temp & 0 \\
\ & \ & \ & \ & \temp & 0 \\ \cline{1-6}
0 & 0 & 0 & 0 & \temp & 0 
\end{array}\right)\;,
\end{align}
\begin{align}
\overline{Q}^{1'} &= \left(\begin{array}{cccccc}
\ & \ & \ & \ & \temp & 0\\ 
\ & \ & \ & \ & \temp & 0 \\
\ & \ & 0_4 & \ & \temp & 0 \\
\ & \ & \ & \ & \temp & 0 \\ \cline{1-6}
0 & 0 & 0 & -1 & \temp & 0 
\end{array}\right)\;,\ \ \ \  
\overline{Q}^{2'} = \left(\begin{array}{cccccc}
\ & \ & \ & \ & \temp & 0\\ 
\ & \ & \ & \ & \temp & 0 \\
\ & \ & 0_4 & \ & \temp & 0 \\
\ & \ & \ & \ & \temp & 0 \\ \cline{1-6}
0 & 0 & 1 & 0 & \temp & 0 
\end{array}\right)\;,\nonumber
\end{align}
\begin{align}\nonumber
S^1 &= \left(\begin{array}{cccccc}
\ & \ & \ & \ & \temp & 0\\ 
\ & \ & \ & \ & \temp & 0 \\
\ & \ & 0_4 & \ & \temp & 0 \\
\ & \ & \ & \ & \temp & 0 \\ \cline{1-6}
1 & 0 & 0 & 0 & \temp & 0 
\end{array}\right)\;,\ \ \ \  
{S}^{2} = \left(\begin{array}{cccccc}
\ & \ & \ & \ & \temp & 0\\ 
\ & \ & \ & \ & \temp & 0 \\
\ & \ & 0_4 & \ & \temp & 0 \\
\ & \ & \ & \ & \temp & 0 \\ \cline{1-6}
0 & 1 & 0 & 0 & \temp & 0 
\end{array}\right)\;,
\end{align}
\begin{align}\nonumber
\overline{S}{}^{1'} &= \left(\begin{array}{cccccc}
\ & \ & \ & \ & \temp & 0\\ 
\ & \ & \ & \ & \temp & 0 \\
\ & \ & 0_4 & \ & \temp & 1 \\
\ & \ & \ & \ & \temp & 0 \\ \cline{1-6}
0 & 0 & 0 & 0 & \temp & 0 
\end{array}\right)\;,\ \ \ \  
\overline{S}{}^{2'} = \left(\begin{array}{cccccc}
\ & \ & \ & \ & \temp & 0\\ 
\ & \ & \ & \ & \temp & 0 \\
\ & \ & 0_4 & \ & \temp & 0 \\
\ & \ & \ & \ & \temp & 1 \\ \cline{1-6}
0 & 0 & 0 & 0 & \temp & 0 
\end{array}\right)\;,\quad 
A = \frac{2i}{3} \left(\begin{array}{cccccc}
\ & \ & \ & \ & \temp & 0\\ 
\ & \ & \ & \ & \temp & 0 \\
\ & \ & \tfrac{1}{4}\,\mathbb{I}_4 & \ & \temp & 0 \\
\ & \ & \ & \ & \temp & 0 \\ \cline{1-6}
0 & 0 & 0 & 0 & \temp & 1 
\label{aagene}
\end{array}\right)\;.
\end{align}
With this choice of basis, we obtain the following (anti)commutation relations
between the (super)generators:
\begin{align}
	[P_a, P_b] &= 0 \;,&& [J_{ab}, P_c] = i\,\eta_{ac}\,P_b-i\,\eta_{bc}\,P_a \;,
 \nonumber\\
	[J_{ab}, J_{cd}] &=  i\,\eta_{ac}\,J_{bd}- i\,\eta_{ad}\,J_{bc} 
    + i\,\eta_{bd}\,J_{ac} -i\,\eta_{bc}\,J_{ad} \;,
    \nonumber\\
	[D,P_a] &= P_a \;,
 \nonumber\\
	[D, K_a] &= - K_a \;, && [P_a, K_b] = -\tfrac{1}{8}\,(J_{ab} + i\,\eta_{ab}\,D) \;,
 \nonumber\\
	[D, J_{ab}] &= 0 \;,
	&& [J_{ab}, K_c] = i\eta_{ac}\,K_b-i\eta_{bc}\,K_a  \;,
 \nonumber\\
	[J_{ab}, Q^A] &= i{(\sigma_{ab})}^A{}_B \,Q^B \;,&& 
    [J_{ab}, {\overline{Q}}^{A'}] = -i{(\tilde{\sigma}_{ab})}_{B'}{}^{A'}\, 
    \overline{Q}^{B'} \;,
    \nonumber\\
	[P_a, Q_\alpha] &= 0 \;,&& [P_a, \overline{Q}^{\alpha'}] = 0 \;,
 \nonumber\\
	\{Q^A, Q^B\} &= 0 \;,&& \{\overline{Q}^{A'}, \overline{Q}^{B'}\} = 0 \;,
 \nonumber\\
	\{Q^{A}, \overline{Q}^{A'}\} &= 2\,{(\sigma^a)}^{AA'}P_a \;,
 \nonumber\\
	[D,Q^A] &= \tfrac{1}{2}\,Q^A  \;,&& 
    [D,\overline{Q}^{A'}] = \tfrac{1}{2}\,\overline{Q}^{A'} \;,
    \nonumber\\
	[K_a, Q^A] &= -\tfrac{1}{4}\,(\sigma_a)^{AA'}\,\overline{S}_{A'} \;,&& 
    [K_a, \overline{Q}^{A'}] = \tfrac{1}{4}\,(\sigma_a)^{AA'}\,S_A \;,
    \nonumber\\
	[J_{ab}, S^A] &= i{(\sigma_{ab})}^A{}_B \, S^B \;,
 && [J_{ab}, \overline{S}^{A'}] 
    = -i{(\tilde{\sigma}_{ab})}_{B'}{}^{A'}\, \overline{S}^{B'}  \;,
    \nonumber\\
   \{S^A,\overline{S}{}^{A'}\} &= 2\,(\sigma^a)^{AA'}\,K_a \;,
   \nonumber\\
	[P_a, S^A] & = -\,\tfrac{1}{4}\,
                      (\sigma_a)^{AA'}\,\overline{Q}_{A'} \;,&&
    [P_a, \overline{S}_{A'}] = -\,\tfrac{1}{4}\,
                      (\tilde\sigma_a)_{A'A}\,{Q}^{A}\;,
    \nonumber\\
	[D, S^A] &= -\tfrac{1}{2}\,S^A \;,&& 
    [D, \overline{S}^{A'}] = - \tfrac{1}{2}\,\overline{S}^{A'}\;,
    \nonumber\\
    \{S^A, Q_B\} 
    &= 
    \tfrac{i}{2}\,(\sigma^{ab})^A{}_B\,J_{ab}+\tfrac{3i}{2}\,\delta_B^A\,A 
    -\tfrac{1}{2}\,\delta_B^A\,D \;,& 
    \nonumber\\
   \{\overline{S}^{A'},\overline{Q}_{B'}\} &= 
    \tfrac{i}{2}\,(\tilde\sigma^{ab})_{B'}{}^{A'}\,J_{ab}
    +\tfrac{3i}{2}\,\delta_{B'}^{A'}\,A 
   +\tfrac{1}{2}\,\delta_{B'}^{A'}\, D\;,
 \nonumber\\
	[A, Q^A] &= -\tfrac{i}{2}\,Q^A \;,&& 
    [A, \overline{Q}^{A'}] = \tfrac{i}{2} \,\overline{Q}^{A'} \;,
    \nonumber\\
	[A, S^A] &= \tfrac{i}{2}\,S^A\;,~ 
    && [A, \overline{S}^{A'}] = -\tfrac{i}{2}\, \overline{S}^{A'}\;.
    \nonumber
\end{align}
As a vector space, the superalgebra $\mathfrak{su}(2,2|1)$ can be 
decomposed through a gradation associated with the eigenvalues of 
the adjoint action of the dilation operator:
\begin{equation}\begin{array}{ccccccccccc}
     \mathfrak{su}(2,2|1) &=&
       ~\mathfrak{g}_{-1} & \oplus & \mathfrak{g}_{-\frac{1}{2}} & \oplus & 
       \mathfrak{g}_{0} & \oplus& \mathfrak{g}_{\frac{1}{2}} & \oplus & \mathfrak{g}_{1}\qquad , \\
       \nonumber \\
      ~\text{Generators:} && K && S && J,A,D && Q && P\qquad .
    \end{array} 
\end{equation}

\section{Superconformal boundary of AdS superspace}
\label{app:AdS}

As we recall in section \ref{sec:Minkowski superspace}, 
(complexified) conformally compact Minkowski superspace 
$\overline{M}^{4|4\mathcal{N}}_{\IC}$ is obtained as the flag supermanifold 
$F(2|0, 2|\mathcal{N}, \IC^{4|\mathcal{N}})$ and is isomorphic to the homogeneous 
space \eqref{Homogeneous space: complex compactified Minkowski} for the (complexified) 
superconformal group $\text{SL}(4|\mathcal{N})$. 
The Lorentzian reality condition is obtained by introducing a hermitian metric 
$h_{\bar{\hat\alpha}\hat\beta}$ and restricting to totally null planes; 
real Lorentzian conformally compact Minkowski superspace $\overline{M}^{4|4\mathcal{N}}$ 
is then isomorphic to the homogeneous space 
\eqref{Homogeneous space: real compactified Minkowski} for the (real Lorentzian) 
superconformal group  $\text{SU}(2,2|\mathcal{N})$. 

Complexified AdS superspace 
$AdS^{4|4\mathcal{N}}_{\IC} \subset \overline{M}^{4|4\mathcal{N}}_{\IC}$ 
and its real Lorentzian counterpart 
$AdS^{4|4\mathcal{N}} \subset \overline{M}^{4|4\mathcal{N}}$ 
are then obtained by reducing the superconformal group to the respective 
group of isometries $\text{Sp}(4|\mathcal{N},\IC)$ and  $\text{OSp}(\mathcal{N}|4)$  
(see figure \ref{Appdx Diagram: group reduction}).
\begin{figure} 
    \begin{equation*}
    \begin{array}{ccc}
      \text{SL}(4|\mathcal{N}, \IC)    &  \xrightarrow{\text{reality condition}} & \text{SU}(2,2|\mathcal{N}) \\[0.4 em] \hspace{2 em} \left\downarrow \begin{array}{c}
           {} \\
            \Lambda <0\\
            {}
      \end{array}\right. && \hspace{2 em} \left\downarrow \begin{array}{c}
           {} \\
            \Lambda <0\\
            {}
      \end{array}\right.\\
        \text{Sp}(4|\mathcal{N}, \IC) &\xrightarrow{\text{reality condition}}  & \text{OSp}(\mathcal{N}|4)
    \end{array}
\end{equation*}
\caption{Summary of the group reductions, horizontal arrows correspond to imposing Lorentzian reality condition while vertical arrows correspond to breaking of superconformal invariance.}\label{Appdx Diagram: group reduction}
\end{figure}
We here would like to briefly recall the corresponding realizations in terms 
of Grasmmanians of $\IC^{4|4\mathcal{N}}$: In order to break superconformal 
invariance one introduces the infinity supertwistors (see e.g. \cite{mason_twistor_2009,adamo_conformal_2014}),
\begin{align}\label{appdx: infinity twistor of AdS}
      I^{\hat\alpha\hat\beta} &= \begin{pmatrix}
        \epsilon^{AB}  & 0 & 0 \\
        0 & \Lambda \epsilon_{A'B'} & 0 \\
        0 & 0 & \sqrt{|\Lambda|} \delta_{IJ}
    \end{pmatrix}, &  I_{\hat\alpha\hat\beta} &= \begin{pmatrix}
        \Lambda \epsilon_{AB}  & 0 & 0 \\
        0 & \epsilon^{A'B'} & 0 \\
        0 & 0 & \sqrt{|\Lambda|} \delta^{IJ}
    \end{pmatrix},
\end{align}
that satisfy 
$I^{\hat\alpha\hat\gamma} I_{\hat\gamma\hat\beta} = |\Lambda| \delta^{\hat\alpha}{}_{\hat\beta}$ 
and coincide with \eqref{eqn: infinity bitwistors} in the limit where $\Lambda$ vanishes. 
For $\Lambda \neq 0$, 
these are skew-symmetric and invertible; as such these define a 
symplectic structure on $\IC^{4|4\mathcal{N}}$. The requirement to preserve this 
symplectic form then realises the left hand side reduction of supergroup in Figure 
\ref{Appdx Diagram: group reduction}. (Complexified) conformally compact Minkowski 
superspace $\overline{M}^{4|4\mathcal{N}}_{\IC} = F(2|0, 2|\mathcal{N}, \IC^{4|\mathcal{N}})$ 
then decompose according to the following: 
if $(Z^{\hat\alpha a}, \tilde{Z}_{\hat\alpha}{}^a)$ 
is a point of $\overline{M}^{4|4\mathcal{N}}_{\IC}$ one can restrict the symplectic form 
\eqref{appdx: infinity twistor of AdS} to the corresponding plane
of $F(2|0, 2|\mathcal{N}, \IC^{4|\mathcal{N}})$:
\begin{align*}
    \sigma \epsilon^{ab} &:= I_{\hat\alpha\hat\beta} Z^{\hat\alpha a} Z^{\hat\beta b}\;, & 
    \tilde \sigma \epsilon_{ab} &:= I^{\hat\alpha\hat\beta} \tilde Z_{\hat\alpha}{}^a 
    \tilde Z_{\hat\beta}{}^b\;. 
\end{align*}
Because of projective invariance the actual values of $\sigma$ and $\tilde\sigma$ are 
irrelevant; the only alternative to consider is whether or not they vanish, i.e., 
whether or not the induced skew symmetric bitwistors are invertible. 
Points of 
$ AdS^{4|4\mathcal{N}}_{\IC}$ corresponds to the situation where both induced symplectic 
structure are non degenerate while points of the superconformal boundary 
$\overline{M}^{3|3\mathcal{N}}$ correspond to planes which are Lagrangian submanifolds:
\begin{align}
    (Z^{\hat\alpha a}, \widetilde{Z}_{\hat\alpha}{}^a)  &\in AdS^{4|4\mathcal{N}}_{\IC}  & \Leftrightarrow&& &\sigma, \tilde \sigma  \in \text{GL}(1|0), \\
    (Z^{\hat\alpha a}, \widetilde{Z}_{\hat\alpha}{}^a) &\in \overline{M}^{3|3\mathcal{N}} & \Leftrightarrow&& &\sigma = \tilde \sigma = 0\;.
\end{align}

One then obtains the real Lorentzian AdS superspace and boundary by imposing the reality 
condition 
$\widetilde{Z}_{\hat\alpha}{}^a = 
\overline{Z}^{\overline{\hat\beta} a}\,h_{\overline{\hat\beta}\hat\alpha} $. 
It is here useful to introduce the map 
$J^{\hat\alpha}{}_{\bar{\hat\beta}} := I^{\hat\alpha\hat\gamma}
h_{\bar{\hat\beta}\hat\gamma}\,$.
It satisfies 
$J^{\hat\alpha}{}_{\bar{\hat\gamma}}\,\overline{J}^{\,\bar{\hat\gamma}}{}_{\hat\beta}=$
$|\Lambda| \delta^{\hat\alpha}{}_{\hat\beta}$ 
and plays the role of a real structure on $\IC^{4|4\mathcal{N}}$. 
Real twistors then correspond to the eigenspace
\begin{align*}
    Z^{\hat\alpha} &\in \IR^{4|4\mathcal{N}}  & \Leftrightarrow && Z^{\hat\alpha} &
    = \frac{1}{\sqrt{|\Lambda|}}J^{\hat\alpha}{}_{\overline{\hat\beta}} \;\overline{Z}^{\overline{\hat\beta}}.
\end{align*}
With this definition, one obtains another characterisation of the real Lorentzian 
superconformal boundary $\overline{M}^{3|3\mathcal{N}}$ as planes in $\IC^{4|\mathcal{N}}$ 
which are both real and Lagrangian (see e.g. \cite{kuzenko_conformally_2012}). 
Finally, since 
$\text{OSp}(\mathcal{N}|4, \IR)$ is the subgroup of $\text{Sp}(4|\mathcal{N},\IC)$ 
stabilizing the real structure $J^{\hat\alpha}{}_{\bar{\hat\beta}}\,$, 
it naturally acts 
on both $AdS^{4|4\mathcal{N}}$ and $\overline{M}^{3|3\mathcal{N}}$. 
One can prove that this action is transitive and that one recovers in this way 
the isomorphisms \eqref{Homogeneous space: intro, super AdS}.


\begin{thebibliography}{10}

\bibitem{penrose_asymptotic_1963}
R.~Penrose, ``Asymptotic {{Properties}} of {{Fields}} and {{Space-Times}},''
  \href{http://dx.doi.org/10.1103/PhysRevLett.10.66}{{\em Physical Review
  Letters} {\bf 10} (1963) no.~2, 66--68}.

\bibitem{geroch_asymptotic_1977}
R.~Geroch, \href{http://dx.doi.org/10.1007/978-1-4684-2343-3_1}{``Asymptotic
  {{Structure}} of {{Space-Time}},''} in {\em Asymptotic {{Structure}} of
  {{Space-Time}}}, pp.~1--105.
\newblock {Springer US}, {Boston, MA}, 1977.

\bibitem{wald_general_1984}
R.~M. Wald, {\em General {{Relativity}}}.
\newblock {The university of Chicago Press}, 1984.

\bibitem{frauendiener_conformal_2004}
J.~Frauendiener, ``Conformal {{Infinity}},''
  \href{http://dx.doi.org/10.12942/lrr-2004-1}{{\em Living Reviews in
  Relativity} {\bf 7} (2004) no.~1, 1}.

\bibitem{ashtekar_geometry_2015}
A.~Ashtekar, ``Geometry and physics of null infinity,''
  \href{http://dx.doi.org/10.4310/SDG.2015.v20.n1.a5}{{\em Surveys in
  Differential Geometry} {\bf 20} (2015) no.~1, 99--122}.

\bibitem{bondi_gravitational_1962}
H.~Bondi, M.~G.~J. {Van der Burg}, and A.~W.~K. Metzner, ``Gravitational waves
  in general relativity, {{VII}}. {{Waves}} from axi-symmetric isolated
  system,'' \href{http://dx.doi.org/10.1098/rspa.1962.0161}{{\em Proceedings of
  the Royal Society of London. Series A. Mathematical and Physical Sciences}
  {\bf 269} (1962) no.~1336, 21--52}.

\bibitem{sachs_gravitational_1962}
R.~K. Sachs, ``Gravitational waves in general relativity {{VIII}}. {{Waves}} in
  asymptotically flat space-time,''
  \href{http://dx.doi.org/10.1098/rspa.1962.0206}{{\em Proceedings of the Royal
  Society of London. Series A. Mathematical and Physical Sciences} {\bf 270}
  (1962) no.~1340, 103--126}.

\bibitem{witten_anti_1998}
E.~Witten, ``Anti de {{Sitter}} space and holography,''
  \href{http://dx.doi.org/10.4310/ATMP.1998.v2.n2.a2}{{\em Advances in
  Theoretical and Mathematical Physics} {\bf 2} (1998) no.~2, 253--291}.

\bibitem{maldacena_large_1998}
J.~Maldacena, ``The large {{N}} limit of superconformal field theories and
  supergravity,'' \href{http://dx.doi.org/10.4310/ATMP.1998.v2.n2.a1}{{\em
  Advances in Theoretical and Mathematical Physics} {\bf 2} (1998) no.~2,
  231--252}.

\bibitem{manin_gauge_1988}
Y.~I. Manin, \href{http://dx.doi.org/10.1007/978-3-662-07386-5}{{\em Gauge
  {{Field Theory}} and {{Complex Geometry}}}}.
\newblock Grundlehren Der Mathematischen {{Wissenschaften}}. {Springer-Verlag},
  {Berlin Heidelberg}, 2~ed., 1988.

\bibitem{ferber_supertwistors_1978}
A.~Ferber, ``Supertwistors and conformal supersymmetry,''
  \href{http://dx.doi.org/10.1016/0550-3213(78)90257-2}{{\em Nuclear Physics B}
  {\bf 132} (1978) no.~1, 55--64}.

\bibitem{castellani_supergravity_1991}
L.~Castellani, R.~D'Auria, and P.~Fre, {\em Supergravity and {{Superstrings}} a
  {{Geometric Perspective}}}.
\newblock {World Scientific Pub Co Inc}, {Singapore ; Teaneck, N.J}, 1991.

\bibitem{buchbinder_ideas_1998}
I.~L. Buchbinder and S.~M. Kuzenko, {\em Ideas and {{Methods}} of
  {{Supersymmetry}} and {{Supergravity}}: {{Or}} a {{Walk Through
  Superspace}}}.
\newblock {CRC Press}, {Bristol ; Philadelphia}, 1st edition~ed., 1998.

\bibitem{galperin_harmonic_2001}
A.~S. Galperin, E.~A. Ivanov, V.~I. Ogievetsky, and E.~S. Sokatchev,
  \href{http://dx.doi.org/10.1017/CBO9780511535109}{{\em Harmonic
  {{Superspace}}}}.
\newblock Cambridge {{Monographs}} on {{Mathematical Physics}}. {Cambridge
  University Press}, {Cambridge}, 2001.

\bibitem{figueroa-ofarrill_kinematical_2019}
J.~{Figueroa-O'Farrill} and R.~Grassie, ``Kinematical superspaces,''
  \href{http://dx.doi.org/10.1007/JHEP11(2019)008}{{\em Journal of High Energy
  Physics} {\bf 2019} (2019) no.~11, 8}.

\bibitem{kuzenko_supertwistor_2021}
S.~M. Kuzenko and G.~{Tartaglino-Mazzucchelli}, ``Supertwistor realisations of
  ads superspaces,'' {\em arXiv:2108.03907 [hep-th, physics:math-ph]} (2021)  ,
  \href{http://arxiv.org/abs/2108.03907}{{\tt\color{arXiv} arxiv:2108.03907
  [hep-th, physics:math-ph]}}.

\bibitem{koning_embedding_2023}
N.~E. Koning, S.~M. Kuzenko, and E.~S.~N. Raptakis, ``Embedding formalism for
  {{N-extended AdS}} superspace in four dimensions,''
  \href{http://arxiv.org/abs/2308.04135}{{\tt\color{arXiv} arxiv:2308.04135
  [hep-th, physics:math-ph]}}.

\bibitem{kuzenko_embedding_2023}
S.~M. Kuzenko and K.~Turner, ``Embedding formalism for (p, q) {{AdS}}
  superspaces in three dimensions,''
  \href{http://dx.doi.org/10.1007/JHEP06(2023)142}{{\em Journal of High Energy
  Physics} {\bf 2023} (2023) no.~6, 142}.

\bibitem{kotrla_supertwistors_1985}
M.~Kotrla and J.~Niederle, ``Supertwistors and superspace,''
  \href{http://dx.doi.org/10.1007/BF01595531}{{\em Czechoslovak Journal of
  Physics B} {\bf 35} (1985) no.~6, 602--620}.

\bibitem{lukierski_general_1988}
J.~Lukierski and A.~Nowicki, ``General superspaces from supertwistors,''
  \href{http://dx.doi.org/10.1016/0370-2693(88)90903-3}{{\em Physics Letters B}
  {\bf 211} (1988) no.~3, 276--280}.

\bibitem{howe_harmonic_1994}
P.~S. Howe and M.~I. Leeming, ``Harmonic superspaces in low dimensions,''
  \href{http://dx.doi.org/10.1088/0264-9381/11/12/004}{{\em Classical and
  Quantum Gravity} {\bf 11} (1994) no.~12, 2843}.

\bibitem{hartwell_n_1995}
G.~Hartwell and P.~Howe, ``({{N}}, p, q) harmonic superspace,''
  \href{http://dx.doi.org/10.1142/S0217751X95001820}{{\em International Journal
  of Modern Physics A} {\bf 10} (1995) no.~27, 3901--3919}.

\bibitem{howe_superspace_1995}
P.~S. Howe and G.~G. Hartwell, ``A superspace survey,''
  \href{http://dx.doi.org/10.1088/0264-9381/12/8/005}{{\em Classical and
  Quantum Gravity} {\bf 12} (1995) no.~8, 1823}.

\bibitem{kuzenko_compactified_2006}
S.~M. Kuzenko, ``On compactified harmonic/projective superspace, {{5D}}
  superconformal theories, and all that,''
  \href{http://dx.doi.org/10.1016/j.nuclphysb.2006.03.019}{{\em Nuclear Physics
  B} {\bf 745} (2006) no.~3, 176--207}.

\bibitem{wolf_supertwistor_2006}
M.~Wolf, {\em On Supertwistor Geometry and Integrability in Super Gauge
  Theory}.
\newblock PhD thesis, 2006.
\newblock \href{http://arxiv.org/abs/hep-th/0611013}{{\tt\color{arXiv}
  arxiv:hep-th/0611013}}.

\bibitem{boels_supersymmetric_2007}
R.~Boels, L.~Mason, and D.~Skinner, ``Supersymmetric gauge theories in twistor
  space,'' \href{http://dx.doi.org/10.1088/1126-6708/2007/02/014}{{\em Journal
  of High Energy Physics} {\bf 2007} (2007) no.~02, 014}.

\bibitem{wolf_first_2010}
M.~Wolf, ``A first course on twistors, integrability and gluon scattering
  amplitudes,'' \href{http://dx.doi.org/10.1088/1751-8113/43/39/393001}{{\em
  Journal of Physics A: Mathematical and Theoretical} {\bf 43} (2010) no.~39,
  393001}.

\bibitem{kuzenko_off-shell_2011}
S.~M. Kuzenko, J.-H. Park, G.~{Tartaglino-Mazzucchelli}, and R.~{von Unge},
  ``Off-shell superconformal nonlinear sigma-models in three dimensions,''
  \href{http://dx.doi.org/10.1007/JHEP01(2011)146}{{\em Journal of High Energy
  Physics} {\bf 2011} (2011) no.~1, 146}.

\bibitem{kuzenko_conformally_2012}
S.~M. Kuzenko, ``Conformally compactified {{Minkowski}} superspaces
  revisited,'' \href{http://dx.doi.org/10.1007/JHEP10(2012)135}{{\em Journal of
  High Energy Physics} {\bf 2012} (2012) no.~10, 135}.

\bibitem{adamo_twistor_2013}
T.~Adamo, {\em Twistor Actions for Gauge Theory and Gravity}.
\newblock PhD thesis, 2013.
\newblock \href{http://arxiv.org/abs/1308.2820}{{\tt\color{arXiv}
  arxiv:1308.2820}}.

\bibitem{buchbinder_superconformal_2015}
E.~I. Buchbinder, S.~M. Kuzenko, and I.~B. Samsonov, ``Superconformal field
  theory in three dimensions: Correlation functions of conserved currents,''
  \href{http://dx.doi.org/10.1007/JHEP06(2015)138}{{\em Journal of High Energy
  Physics} {\bf 2015} (2015) no.~6, 138}.

\bibitem{adamo_perturbative_2014}
T.~Adamo, E.~Casali, and D.~Skinner, ``Perturbative gravity at null infinity,''
  \href{http://dx.doi.org/10.1088/0264-9381/31/22/225008}{{\em Classical and
  Quantum Gravity} {\bf 31} (2014) no.~22, 225008}.

\bibitem{Witten:1985nt}
E.~Witten, ``{Twistor - Like Transform in Ten-Dimensions},''
  \href{http://dx.doi.org/10.1016/0550-3213(86)90090-8}{{\em Nucl. Phys. B}
  {\bf 266} (1986)  245--264}.

\bibitem{Mason:2013sva}
L.~Mason and D.~Skinner, ``{Ambitwistor strings and the scattering
  equations},'' \href{http://dx.doi.org/10.1007/JHEP07(2014)048}{{\em JHEP}
  {\bf 07} (2014)  048},
  \href{http://arxiv.org/abs/1311.2564}{{\tt\color{arXiv} arXiv:1311.2564
  [hep-th]}}.

\bibitem{dewitt_supermanifolds_1992}
B.~DeWitt, \href{http://dx.doi.org/10.1017/CBO9780511564000}{{\em
  Supermanifolds}}.
\newblock Cambridge {Monographs} on {Mathematical} {Physics}. Cambridge
  University Press, Cambridge, 2~ed., 1992.
\newblock
  \url{https://www.cambridge.org/core/books/supermanifolds/B9914F457D773C82B850246FAA2B74EE}.

\bibitem{rogers2007supermanifolds}
A.~Rogers, {\em Supermanifolds: theory and applications}.
\newblock World Scientific, 2007.

\bibitem{tuynman2004supermanifolds}
G.~M. Tuynman, {\em Supermanifolds and supergroups: basic theory}, vol.~570.
\newblock Springer Science \& Business Media, 2004.

\bibitem{mason_twistor_2009}
L.~J. Mason and M.~Wolf, ``Twistor {{Actions}} for {{Self-Dual
  Supergravities}},'' \href{http://dx.doi.org/10.1007/s00220-009-0732-5}{{\em
  Communications in Mathematical Physics} {\bf 288} (2009) no.~1, 97--123}.

\bibitem{adamo_conformal_2014}
T.~Adamo and L.~Mason, ``Conformal and {{Einstein}} gravity from twistor
  actions,'' \href{http://dx.doi.org/10.1088/0264-9381/31/4/045014}{{\em
  Classical and Quantum Gravity} {\bf 31} (2014) no.~4, 045014}.

\end{thebibliography}
\providecommand{\href}[2]{#2}\begingroup\raggedright\endgroup

\end{document}